\documentstyle[prd,aps,psfig]{revtex}
\tighten
\begin{document}
\draft
\title{Quintessence and Gravitational Waves}
\author{Alain Riazuelo}
\address{D\'epartement d'Astrophysique Relativiste et de Cosmologie,\\
         CNRS-UMR 8629, Observatoire de Paris,
         F-92195 Meudon  (France)\\
         e-mail: Alain.Riazuelo@obspm.fr}
\author{Jean-Philippe Uzan}
\address{Laboratoire de Physique Th\'eorique, CNRS-UMR 8627,\\
         Universit\'e Paris XI, B\^at. 210, 
         F-91405 Orsay Cedex (France) and \\
         D\'epartement de Physique Th\'eorique, Universit\'e de
         Gen\`eve,\\
         24 quai E.~Ansermet, CH-1211 Gen\`eve 4 (Switzerland)\\
         e-mail: Jean-Philippe.Uzan@th.u-psud.fr}
\date{20 may 2000}
\maketitle


\newcommand{\ETAL}{{\it et al.}}
\newcommand{\IE}{{i.e.}}
\newcommand{\EG}{{e.g.}}

\newcommand{\PL}{{\rm Pl}}
\newcommand{\GW}{{\rm GW}}
\newcommand{\EFF}{{\rm eff}}
\newcommand{\FLUID}{{\rm fluid}}
\newcommand{\LSS}{{\rm LSS}}

\newcommand{\ICE}{{(IC1)}}
\newcommand{\ICK}{{(IC2)}}

\newcommand{\Hconf}{{\cal H}}
\newcommand{\ddd}{{\rm d}}

\newcommand{\FIG}[1]{Fig.~{#1}}
\newcommand{\FIGS}[1]{Figs.~{#1}}
\newcommand{\FFIG}[1]{Figure~{#1}}
\newcommand{\FFIGS}[1]{Figures~{#1}}

\newcommand{\EQN}[1]{Eq.~{#1}}
\newcommand{\EQNS}[1]{Eqns.~{#1}}
\newcommand{\EEQN}[1]{Equation~{#1}}
\newcommand{\EEQNS}[1]{Equations~{#1}}


\begin{abstract}
We investigate some aspects of quintessence models with a
non-minimally coupled scalar field and in particular we show that it
can behave as a component of matter with $-3 \lesssim P/\rho \lesssim
0$. We study the properties of gravitational waves in this class of
models and discuss their energy spectrum and the cosmic microwave
background anisotropies they induce. We also show that gravitational
waves are damped by the anisotropic stress of the radiation and that
their energy spectrum may help to distinguish between inverse power
law potential and supergravity motivated potential. We finish by a
discussion on the constraints arising from their density parameter
$\Omega_\GW$.
\end{abstract}
\pacs{{\bf Preprint :} DARC 00-01, LPT-ORSAY 00/41, UGVA-DPT 00/04-1076}
\vskip2pc


\section{Introduction}\label{par1}

Recent astrophysical and cosmological observations such as the
luminosity distance-redshift relation for the supernovae
type~Ia~\cite{sn1a_1,sn1a_2,sn1a_3}, the recent observations of the
cosmic microwave background temperature anisotropies~\cite{boomerang},
gravitional lensing~\cite{lens} and velocity fields~\cite{roman} tend
to indicate that a large fraction of the matter of the universe today
is composed of matter with negative pressure (see
\EG~\cite{indication} for a comparison of the different observations).
Recent analyses~\cite{efsta,perlmutter99} seem to indicate that the
energy density $\rho$ and the pressure $P$ of this fluid satisfies
\begin{equation}
-1\leq P/\rho \leq -0.6,
\end{equation}
which is compatible with a cosmological constant $\Lambda$ for which
$P/\rho=-1$ (see also~\cite{phantom} for arguments in favour of
$P/\rho<-1$). A typical value of $\Omega_\Lambda\simeq 0.7$ for its
energy density in units of the critical density of the universe
corresponds to an energy scale of order
$5\times10^{-47}\,\hbox{GeV}^4$ which is very far from what is
expected from high energy physics; this is the well known cosmological
constant problem~\cite{weinberg}. To circumvent this problem different
solutions have been proposed starting from the idea of a dynamical
cosmological constant~\cite{coble} to lead to the class of models
known as {\it quintessence}~\cite{caldwell}, where a spatially
homogeneous scalar field $\phi$ is rolling down a potential decreasing
when $\phi$ tends to infinity. An example of such a potential which
has been widely studied is the inverse power law potential. It can be
obtained from some high energy physics models, \EG{} where
supersymmetry is broken through fermion
condensates~\cite{binetruy}. Recently, it has been argued~\cite{brax}
that supergravity has to be taken into account since today one expects
the scalar field to be of order of the Planck mass $M_\PL$ and
corrections to the potential appear at this energy. This leads to a
better agreement with observations~\cite{brax2}.

An important point about this family of models is the existence of
scaling solutions~\cite{scaling1,scaling2} (refered to as {\it
tracking solutions}), \IE{} such that $\phi$ evolves as the scale
factor of the universe at a given power. These solutions are
attractors of the dynamical system describing the evolution of the
scale factor and of the scalar field. This implies that the present
time behaviour of the field is almost independent of its initial
conditions~\cite{steinhard,steinhard2}. This property allows to
address~\cite{zlatev} (i) the {\it coincidence problem},
\IE{} the fact that $\phi$ starts to dominate today and (ii) the {\it
fine tuning problem}, \IE{} the fact that one does not have to fine
tune the initial condition of the field $\phi$.

One of us extended these models to include a non-minimal coupling
$\xi\bar Rf(\phi)$ between the scalar field and the scalar curvature
$\bar R$~\cite{uzan99}. Such a coupling term appears \EG~when
quantising fields in curved spacetime~\cite{birrel,ford87} and in
multi-dimensional theories~\cite{maeda,acceta,mata}. It was shown that
when $f(\phi)=\phi^2/2$ tracking solutions still exist~\cite{uzan99}
and this result was generalised~\cite{amendola} to any coupling
function $f$ and potential $V$ satisfying $V(\phi)\propto
f^n(\phi)$. However, such a coupling is constrained by the variations
of the constants of nature~\cite{caroll} which fix bounds on
$\xi$~\cite{chiba99}. A way to circumvent this problem is to consider
quintessence models in the framework of scalar-tensor
theories~\cite{bartolo,bertolami} where a double attractor mechanism
can occur, \IE{} of the scalar-tensor theory towards general
relativity and of the scalar field $\phi$ towards its tracking
solution.

Among all the possible observations of cosmology, gravitational waves
give an insight on epochs where there was a variation of the
background dynamics since every such variation affects the shape of
the stochastic graviton background spectrum~\cite{gw1,gw2}. We can
then view our universe as containing a sea of stochastic gravitational
waves from primordial origin, as predicted by most models of structure
formations such as inflation~\cite{gw1,gw2} (see also~\cite{allen94}
for a review) and topological defects scenarios~\cite{vs}. Their
spectrum extends typically from $10^{-18}$~Hz (for wavelengths of
order of the size of the Hubble radius today) to $\simeq 10^{10}$~Hz
(the smallest mode that has been inflated out of the Hubble radius)
and they could be detected or constrained by coming experiments such
as LIGO~\cite{ligo}, VIRGO~\cite{virgo} (at $\simeq 10^{2}$~Hz) and
LISA~\cite{lisa} (at $\simeq 10^{-4}$~Hz). Gravitational waves, which
are perturbations in the metric of the universe have also an effect on
the cosmic microwave (CMB) temperature
anisotropy~\cite{soft1,soft2,soft3,white92,turner93} and
polarisation~\cite{pola} allowing to extract information on their
amplitude from the measure of the CMB anisotropies. For instance,
bounds on the energy density spectrum of these cosmological
gravitational waves in units of the critical density, $\Omega_\GW$,
have been obtained from the CMB~\cite{soft1,soft2,soft3} $$
\left.\frac{\ddd
\Omega_\GW}{\ddd \ln \omega}\right|_{10^{-18}\,\mathrm{Hz}}
\lesssim 10^{-10}.
$$

Gravitational waves are also a very good probe of the conditions in
the early universe since they decouple early in its history and can
help \EG{} testing the initial conditions of $\phi$. An example was
put forward by Giovannini~\cite{giovannini99,giovannini99b} who showed
that in a class of quintessential inflation models~\cite{vilenkin}
there was an era dominated by the scalar field $\phi$ before the
radiation dominated era which implies that a large part of the
gravitational wave energy of order $\Omega_\GW\simeq 10^{-6}$ (about
eight orders of magnitude higher than for standard inflation) was in
the GHz region. This may happen in any scenario where the inflation
ends with a kinetic phase~\cite{ford87,spoko93} or when the dominant
energy condition is violated \cite{gio3}. On the other hand, the CMB
temperature fluctuations give information on the history of the
gravitational waves in between the last scattering surface and today
through the integrated Sachs-Wolfe effect, whereas the polarisation of
the CMB radiation gives mainly information on the gravitational waves
at decoupling. These three observables (energy spectrum, CMB
temperature and polarisation anisotropies) are thus complementary and
we aim to present here a global study of the cosmological properties
of the gravitational waves.

\vspace{0.5cm}

The goals of this article are (i) to study in more details the cosmology
with a non-minimal quintessence field and (ii) to study gravitational
waves in this class of models. In \S\ref{par2} we set up the general
framework and describe the two potentials we shall consider. In
\S\ref{par3} we introduce and define the observable quantities
associated with the gravitational waves: their energy density spectrum
and their imprint on the CMB radiation anisotropies and its
polarisation. In \S\ref{sec_damping}, we point out the general mechanism
of damping by the anisotropic stress of the radiation. In \S\ref{par4}
we discuss the parameters of the problem and investigate the tuning of
the potential parameters; we also describe the evolution of the
background spacetime and show that a non-minimally coupled quintessence
field is a candidate for a ($\omega<-1$)-matter. In \S\ref{par5} we
describe the main properties of the gravitational waves. We finish in
\S\ref{par6} by presenting numerical results and we underline the
complementarity of the different observational quantities.

This work gives a detailed study of the observational effects of
gravitational waves in the framework of quintessence, including some
recent developments, and allowing for non-minimal coupling. This
extends the work on quintessential inflation~\cite{giovannini99} by
including the effects on the CMB. It also extends the studies on
$\Lambda$CDM~\cite{melchiorri99} to quintessence and is, as far as we
know, a more complete study of the effect of gravitational waves on
the CMB polarisation. We hope to show that a joint study of the
gravitational wave detection experiments~\cite{ligo,virgo,lisa} of the
CMB experiments~\cite{boomerang,map,planck} and of the polarisation
experiments~\cite{planck} can lead to a better determination of their
properties.


\section{General framework}
\label{par2}

\subsection{Background spacetime}\label{par2_1}

We consider a universe described by a Friedmann-Lema\^{\i}tre model
with Euclidean spatial sections so that the metric takes the form
\begin{equation}\label{metrique}
\ddd s^2=a^2(\eta)\left[-\ddd\eta^2
                        +\delta_{ij}\ddd x^i \ddd x^j\right]\equiv
     \bar g_{\mu\nu}\ddd x^\mu \ddd x^\nu,
\end{equation}
where $a$ is the scale factor and $\eta$ the conformal time. Greek
indices run from 0 to 3 and latin indices from 1 to 3.

We assume that the matter content of the universe can be described by a
mixure of matter and radiation (mainly baryons, CDM, photons and three
families of massless, non-degenerate neutrinos) and a scalar field
$\phi$ non-minimally coupled to gravity evolving in a potential
$V(\phi)$ that will be described later. The action for this system is
\begin{equation}\label{action}
S=\int \ddd^4x\sqrt{-\bar g}\left[\frac{\bar R}{2\kappa}-\xi\bar R f(\phi)
-\frac{1}{2}\partial_\mu\phi\partial^\mu\phi-V(\phi)+ {\cal L}_{\rm
matter}\right],
\end{equation}
with $\kappa\equiv 8\pi G$, $G$ being the Newton constant, and where
${\cal L}_{\rm matter}$ is the Lagrangian of the ordinary matter which
is uncoupled to the scalar field and $f(\phi)$ is an arbitrary
function of the scalar field that will be specified later. The
action~(\ref{action}) can be rewritten under the interesting form
\begin{equation}\label{action2}
S=\int \ddd^4x\sqrt{-\bar g}\left[\frac{\bar R}{2\kappa_\EFF[\phi]}
  -\frac{1}{2}\partial_\mu\phi\partial^\mu\phi-V(\phi)+
   {\cal L}_{\rm matter}\right],
\end{equation}
with
\begin{equation}
\label{kappa_eff}
\kappa_\EFF[\phi] \equiv \frac{\kappa}{1-2\xi\kappa f(\phi)}.
\end{equation}
The stress-energy tensor of the scalar field is obtained by
varying its Lagrangian [$-\xi\bar R f(\phi)
-\frac{1}{2}\partial_\mu\phi\partial^\mu\phi-V(\phi)$] to get
\begin{eqnarray}\label{Tphi}
T_{\mu\nu}^{(\phi)}
 =   \bar\nabla_\mu\phi\bar\nabla_\nu\phi
   - \frac{1}{2}{\bar g}_{\mu\nu}\bar\nabla_\lambda \phi\bar\nabla^\lambda\phi
   - V(\phi){\bar g}_{\mu\nu}
   +2\xi\left[  {\bar g}_{\mu\nu}\bar\nabla_\lambda\phi\bar\nabla^\lambda\phi
              - \bar\nabla_\mu\phi\bar\nabla_\nu\phi
              - \phi\bar\nabla_\mu\bar\nabla_\nu\phi
              + \phi\Box\phi{\bar g}_{\mu\nu}
              + {\bar G}_{\mu\nu}f(\phi)\right]
\end{eqnarray}
where $\bar G_{\mu\nu}$ is the Einstein tensor of the metric $\bar
g_{\mu\nu}$, $\bar\nabla$ its covariant derivative and
$\Box\equiv\bar\nabla_\mu\bar\nabla^\mu$.

The equations governing the evolution of the background spacetime are
then obtained by varying~(\ref{action}) with respect to $\bar
g_{\mu\nu}$, $\phi$ and the ordinary matter fields to get respectively
the Friedmann equations, the Klein-Gordon equation and the fluid
conservation equation
\begin{eqnarray}
\Hconf^2&=&\frac{\kappa a^2}{3}(\rho+\rho_\phi),\label{eq1}\\
\dot\Hconf-\Hconf^2&=&-\frac{\kappa
a^2}{2}(\rho+P+\rho_\phi+P_\phi),\label{eq2}\\
\ddot \phi
 &+& 2\Hconf\dot\phi
+a^2\frac{\ddd V}{\ddd\phi}
+6\xi(2\Hconf^2+\dot\Hconf)=0,\label{eq3}\\
\dot\rho&=&-3\Hconf(\rho+P).
\end{eqnarray}
A dot denotes a derivative with respect to the conformal time and
$\Hconf \equiv \dot a/a$ is the comoving Hubble parameter. The matter
fluid energy density $\rho$ and pressure $P$ are assumed to satisfy
the equation of state $P=\omega\rho$. The factor $\omega$ varies from
$1/3$ deep in the radiation era to $0$ in the matter era. The scalar
field energy density $\rho_\phi$ and pressure $P_\phi$ are obtained
from its stress-energy tensor (\ref{Tphi}) and are
then explicitely given by
\begin{eqnarray}
\rho_\phi
 & = &   \frac{1}{2}\frac{\dot\phi^{2}}{a^2} + V(\phi)
       + \frac{2\xi}{a^2} \left[  3\Hconf^2 f(\phi)
                                + 3\Hconf \dot f(\phi)\right], \\
P_\phi
 & = &   \frac{1}{2}\frac{\dot\phi^{2}}{a^2} - V(\phi)
       - \frac{2\xi}{a^2}\left[  (2\dot\Hconf+3\Hconf^2)f(\phi)
                               + \Hconf\dot f(\phi)+\ddot f(\phi)\right].
\end{eqnarray}
We stress that the conservation equation derived from~(\ref{Tphi})
reduces to the Klein-Gordon equation~(\ref{eq3}). For each matter
component, $X$ say, we introduce the density parameter $\Omega_X$
defined as
\begin{equation}
\Omega_X\equiv\frac{\kappa a^2\rho_X}{3\Hconf^2}.
\end{equation}

To completely specify the model, we have to fix the potential
$V(\phi)$. Following our previous work~\cite{uzan99} and as discussed
in the introduction we choose it to behave as
\begin{equation}\label{pot}
V(\phi)= \Lambda^4\left(\frac{\Lambda}{\phi}\right)^\alpha
\quad,\quad\alpha>0.
\end{equation}
where $\Lambda$ is an energy scale. As shown in~\cite{uzan99}, such a
potential leads to the existence of tracking solutions whatever the
value of $\xi$ and for which the scalar field behaves as a barotropic
fluid of equation of state (as long as the background fluid dominates)
\begin{equation}\label{eqstatephi}
P_\phi=\omega_\phi\rho_\phi\qquad\hbox{with}\qquad
\omega_\phi = -1 + \frac{\alpha(1+\omega)}{\alpha+2}.
\end{equation}
We also consider another class of potentials arising when one takes
supergravity into account~\cite{brax,brax2} and given by
\begin{equation}\label{potsugra}
\widetilde V(\phi) = \Lambda^4\left(\frac{\Lambda}{\phi}\right)^\alpha
	         \exp(\kappa\phi^2/2)
\quad,\quad\alpha>0.
\end{equation}
The effect of the exponential term is important only at late time so
that the scaling properties of the tracking solution are not affected
during the matter and radiation era. However when the field starts to
dominate it leads to its stabilisation~\cite{brax2} which has an effect
on the effective equation of state of the scalar fluid.

\subsection{Gravitational waves}\label{par2_2}

In this article, we want to focus on the properties of the
gravitational waves which are tensorial perturbations. At linear
order, the metric is expanded as
\begin{equation}
g_{\mu\nu}=\bar g_{\mu\nu}+f_{\mu\nu}
\end{equation}
where $f_{\mu\nu}$ is a transverse traceless (TT) perturbation,
\IE{} satisfying
\begin{equation}\label{TT1}
f_{00}=f_{0i}=0,\quad f_{\mu\nu}\bar
g^{\mu\nu}=0,\quad\bar\nabla_\mu f^{\mu\nu}=0.
\end{equation}
It is also useful to define the perturbation $h_{\mu\nu}$ by
\begin{equation}
f_{\mu\nu}\equiv a^2h_{\mu\nu}
\end{equation}
which, from~(\ref{TT1}), satisfies
\begin{equation}\label{defh}
h_{00}=h_{0i}=0,\quad h_{kl}\delta^{kl}=0,\quad\partial_k
h^{kl}=0.
\end{equation}

The equation of evolution of $h_{kl}$ is obtained by considering the
TT part of the perturbed Einstein equation (see \EG~\cite{kodama})
which leads to
\begin{equation}\label{eqh}
\ddot h_{kl}+2\left[\Hconf-\kappa_\EFF \xi \dot f(\phi) \right]
\dot h_{kl}-\Delta h_{kl}=2\kappa Pa^2\bar\pi_{kl}
\end{equation}
where $\Delta\equiv\partial_i\partial^i$ is the Laplacian and where
the anisotropic stress tensor of the matter, $\bar\pi_{kl}$, is
defined as the tensor component of the matter stress-energy tensor
\begin{equation}
\delta T^i_j\equiv P\bar\pi^i_j,\quad
\bar\pi^k_k=\partial_i\bar\pi^i_k=0.
\end{equation}
The anisotropic stress of the matter fluid is dominated by the
contribution of the neutrinos and of the photons and its form can be
obtained by describing these relativistic fluids by a Boltzmann
equation~\cite{uzan95} (see \S\ref{par3_2} below).


\section{Observational quantities}
\label{par3}

The goal of this section is to define the observable quantities
related to the gravitational waves. We start by reviewing the
computation of the energy density $\rho_\GW$ of a stochastic
background of gravitational waves and finish by describing their
effect on the CMB radiation, namely we present the computation of the
coefficients $C_\ell$ of the development of the angular correlation
function of the CMB temperature anisotropy and polarisation.

\subsection{Gravitational waves energy density}\label{par3_1}

The definition of the gravitational waves stress-energy tensor
$t_{\mu\nu}$ can be found in \EG~\cite{weinberg} in the case of a
Minkowski background spacetime, in \EG~\cite{ford77} in the case of a
Friedmann-Lema\^{\i}tre spacetime and a general discussion can be
found in \EG~\cite{MTW}.

To define the gravitational waves stress-energy tensor, we have to
expand the Einstein-Hilbert action~(\ref{action2}) to second order in
the perturbation $f_{\mu\nu}$, which implies to develop the curvature
scalar $R$ at second order in the perturbations
(see~\cite{weinberg,ford77,MTW}) to get (up to divergence terms and
forgetting the contribution arising from ${\cal L}_{\rm matter}$)
\begin{equation}
\delta^{(2)}S=-\int\frac{1}{4\kappa_\EFF[\phi]}\bar\nabla_\mu
f_{\alpha\beta} \bar\nabla^\mu f^{\alpha\beta}\sqrt{-\bar g}\ddd^4x.
\end{equation}
This expression is valid whatever the background metric as long as
$f_{\mu\nu}$ is a transverse traceless perturbation. Note that
contrarily to the ``standard'' situation, $\kappa$ now depends on
$\phi$ because of the non-minimal coupling with the scalar field.
Using the fact that $\bar\nabla_\mu f_{\alpha\beta}=a^2\partial_\mu
h_{\alpha\beta}$ we can rewrite the previous expression as
\begin{equation}\label{24}
\delta^{(2)}S=-\int\frac{1}{4\kappa_\EFF[\phi]}\partial_\mu
h_{kl}
\partial^\mu h^{kl}\sqrt{-\bar g}\ddd^4x,
\end{equation}
which assumes a Friedmann-Lema\^{\i}tre background and the
decomposition~(\ref{defh}). Now, we decompose $h_{kl}$ on its two
polarisations as
\begin{equation}
h_{kl}=\sum_{\lambda=+,\times}
 h^{(\lambda)}(\eta,{\bf x})\epsilon_{kl}^{(\lambda)}({\bf x}),
\end{equation}
where $\epsilon_{kl}^{(\lambda)}({\bf x})$ is the polarisation tensor
defined as
\begin{equation}\label{defepsilon}
\epsilon_{kl}^{(\lambda)}({\bf x})\equiv\left(e^1_ke^1_l-e^2_ke^2_l\right)\delta^\lambda_\times +
\left(e^1_ke^2_l+e^1_le^2_k\right)\delta^\lambda_+
\end{equation}
for a wave propagating along the direction ${\bf e}_3$ and where
$({\bf e}_1,{\bf e}_2,{\bf e}_3)$ is a local orthonormal basis. Since
this basis and the polarisation tensor satisfy
\begin{equation}
e^a_ie^i_b=\delta^a_b,\qquad
\epsilon_{kl}^{(\lambda)}\epsilon^{kl}_{(\lambda')}=
2\delta^\lambda_{\lambda'},
\end{equation}
we can rewrite the action~(\ref{24}) of the graviton as
\begin{equation}
\delta^{(2)}S=-\sum_\lambda\int\frac{1}{2\kappa_\EFF [\phi]}
\partial_\mu h^{(\lambda)}
\partial^\mu h^{(\lambda)}\sqrt{-\bar g}\ddd^4x,
\end{equation}
which is the action for two massless scalar fields $h_\lambda$
evolving in the background spacetime, as first noticed by
Grishchuk~\cite{gw1,gw2}. By varying this action with respect to the
background metric, we then deduce the stress-energy tensor of the
gravitational waves
\begin{equation}\label{TGW}
t_{\mu\nu}=-\frac{1}{2\kappa_\EFF[\phi]}
\sum_\lambda\left(\partial_\mu h^{(\lambda)}
\partial_\nu h^{(\lambda)}-\bar g_{\mu\nu}\partial_\alpha
h^{(\lambda)}\partial_\alpha h^{(\lambda)}
\right).
\end{equation}
If we decompose $h^{(\lambda)}$ in Fourier modes as
\begin{equation}
h^{(\lambda)}({\bf x},\eta)=\int\frac{\ddd^3{\bf k}}{(2\pi)^{3}}\hat
h^{(\lambda)}({\bf k},\eta) \hbox{e}^{i{\bf k}.{\bf x}},
\end{equation}
we can relate $\hat h^{(\lambda)}({\bf k},\eta)$ to its initial value
$\hat h^{(\lambda)}({\bf k},\eta_{\rm in})$ , \IE{} its value deep in
the radiation era (\EG{} at the end of the inflationary phase) through
the transfer function $T(k,\eta)$ by solving~(\ref{eqh}) to get
\begin{equation}
\hat h^{(\lambda)}({\bf k},\eta)=T(k,\eta)
\hat h^{(\lambda)}({\bf k},\eta_{\rm in}).
\end{equation}
Defining the initial power spectrum of the tensor modes as
\begin{equation}\label{defPT}
\left<\hat h^{(\lambda)}({\bf k},\eta_{\rm in}) 
      \hat h_{(\lambda')}^* ({\bf k'},\eta_{\rm in})\right>
 \equiv k^{-3}P_h(k) \delta({\bf k}-{\bf k'})\delta^\lambda_{\lambda'},
\end{equation}
($\delta$ is the Dirac distribution), we can express the space average
of $t_0^0({\bf x},\eta)$ as
\begin{equation}\label{too}
-\left<t^0_0({\bf x},\eta)\right> = 
\frac{1}{2\kappa_\EFF[\phi]a^2}\sum_\lambda\left<
\partial_i h^{(\lambda)}\partial_j h^{(\lambda)}\delta^{ij}\right>=
\frac{1}{\kappa_\EFF[\phi]}
\int\frac{k^2}{2\pi a^2}P_h(k) T^2(k,\eta) \ddd \ln(k),
\end{equation}
where we used an ergodic hypothesis to replace the space average by an
ensemble average. Now, since $\left<t^0_0({\bf x},\eta)\right>$
oscillates, we define the energy density of the gravitational waves by
taking the average of~(\ref{too}) over $n$ periods.  It follows that
\begin{equation}
\rho_\GW(\eta)=\frac{1}{\kappa_\EFF[\phi]}
\int\frac{k^2}{2\pi a^2}P_h(k)\bar T^2(k,\eta) \ddd \ln(k),
\end{equation}
where $\bar T(k,\eta)$ is the root mean square of $T(k,\eta)$ over $n$
periods which is well defined as long as the amplitude of the wave
varies slowly with respect to its period.

The energy density $\rho_\GW$ and energy density parameter
$\Omega_\GW$ by frequency band are then obtained (after averaging on
several periods of the wave) by
\begin{eqnarray}
\label{drho}
\frac{\ddd\rho_\GW(k,\eta)}{\ddd\ln(k)}
&=&\frac{1}{2\pi^2\kappa_\EFF[\phi]}\left(
\frac{k}{a}\right)^2P_h(k)\bar T^2(k,\eta),\label{11}\\
\label{dOGW}
\frac{\ddd \Omega_\GW(k,\eta)}{\ddd\ln(k)}
&=&\frac{1}{6\pi^2}\left(
\frac{\kappa}{\kappa_\EFF[\phi]} \right)
\left(\frac{k}{\Hconf}\right)^2P_h(k)\bar T^2(k,\eta)\label{12}.
\end{eqnarray}

Let us stress some important points. Since we have to average on
several periods, these expressions are valid only in a ``shortwave
limit'' (see~\cite{MTW} for discussion) where (i) the amplitude of the
perturbation is small and (ii) the wavelength of the wave is small
compared to the typical radius of the background spacetime. In our
case this can be rephrased as $k/\Hconf>1$ which implies that the
expressions~(\ref{11}-\ref{12}) are valid only for modes which are
``subhorizon'' today, \IE{} whose wavelength is smaller than the
Hubble radius today. For such modes the ergodic hypothesis is well
justified. In fact, because of the averaging procedure of the transfer
function, we have to restrict to modes such that $k/\Hconf_0 \gtrsim
60$ if we want to average on about ten periods. Again, we emphasize
that there is an explicit dependence of $\Omega_\GW$ and $\rho_\GW$ on
the scalar field $\phi$ because of the non-minimal coupling and our
expressions reduce to the standard
ones~\cite{weinberg,allen94,white92,turner93,giovannini99,MTW} when
$\xi=0$. We have described the gravitational waves by two stochastic
variables $\hat h^{(\lambda)}$ which can be understood as being the
classical limit of a complete quantum description of the gravitational
waves (see \EG{}~\cite{gw1,gw2,ford77} for details).

Before turning to the effects of the gravitational waves on the CMB,
let us make a comment that will lead us to introduce some new
notations. In the previous analysis we decomposed the metric
perturbation $h_{ij}$ on the basis $\widetilde Q^\lambda_{ij}({\bf
x},{\bf k})
\equiv\epsilon^\lambda_{ij}\hbox{exp}(i{\bf k}.{\bf x})$ of TT
eigenfunctions of the Laplacian, \IE{} such that
\begin{equation}\label{TTeigen}
\Delta \widetilde Q^\lambda_{ij}=-k^2\widetilde
Q^\lambda_{ij}\quad\hbox{with}\quad
\partial^i\widetilde Q^\lambda_{ij}=\delta^{ij}\widetilde Q^\lambda_{ij}=0.
\end{equation}
Such a decomposition is indeed not unique and we could have chosen any
other such basis. In the CMB literature, one often
prefers~\cite{huwhite97} to use the basis $Q^{(\pm2)}_{ij}({\bf
x},{\bf k})$ defined by
\begin{equation}
Q^{(\pm2)}_{ij}\equiv-\sqrt{\frac{3}{8}}(e_1\pm ie_2)_i (e_1\pm ie_2)_j
\hbox{e}^{i{\bf k}.{\bf x}} ,
\end{equation}
the vectors $e_1$ and $e_2$ being defined above~(\ref{defepsilon}). If
we decompose $h_{ij}$ on the latter basis as
\begin{equation}\label{develop2}
h_{ij}=\sum_{m=\pm2}\int\frac{\ddd^3{\bf
k}}{(2\pi)^3}2H^{(m)}Q^{(m)}_{ij}({\bf x},{\bf k}),
\end{equation}
then the two decompositions are related by
\begin{equation}
\hat h^{(\times)}=-\sqrt{\frac{3}{2}}\left[H^{(+2)}+H^{(-2)}\right]
\quad,\quad
\hat h^{(+)}=-\sqrt{\frac{3}{2}}i\left[H^{(+2)}-H^{(-2)}\right].
\end{equation}
The two polarisations $H^{(\pm2)}$ are then solution of~(\ref{eqh})
which reads
\begin{equation}
\label{evoH}
\ddot H^{(m)}+2\left[\Hconf-\kappa_\EFF \xi\dot
f(\phi)\right]\dot H^{(m)}+k^2H^{(m)}=\kappa P a^2\pi^{(m)}
\end{equation}
where $\pi^{(m)}$ is the coefficient of the development of $\delta
T_{ij}$ as in~(\ref{develop2}) so that the transfer functions for
$H^{(m)}$ and $\hat h^{(\lambda)}$ are the same. If we now define the
power spectrum of $H^{(m)}({\bf k},\eta_{\rm in})$ as
\begin{equation}
\left<H^{(m_1)}({\bf k},\eta_{\rm in})H^{(m_2)*}({\bf k'},\eta_{\rm
in})\right>=(2\pi)^3k^{-3}P_T(k)\delta({\bf k}-{\bf
k'})\delta_{m_1,m_2}
\end{equation} one can easily check that
\begin{equation}
P_T(k)=\frac{1}{3}P_h(k)
\end{equation}
and that if the two polarisations $+$ and $\times$ are independent
then $H^{(+2)}$ and $H^{(-2)}$ are also independent. With these
notations, the energy density spectra are given as
\begin{eqnarray}
\frac{\ddd \rho_\GW}{\ddd\ln(k)}
&=&\frac{3}{2\pi^2\kappa}\left(\frac{k}{a}\right)^2P_T(k)\bar
T^2(k,\eta) ,\\ 
\label{dodlnk}
\frac{\ddd \Omega_\GW}{\ddd\ln(k)}
&=&\frac{1}{2\pi^2}\left(\frac{\kappa}{\kappa_\EFF}\right)
\left(\frac{k}{\Hconf}\right)^2P_T(k)\bar T^2(k,\eta).
\end{eqnarray}
Indeed, this does not change the result but we found interesting to
make the link between the notations used in the gravitational waves
literature~\cite{weinberg,allen94,white92,turner93,giovannini99,MTW}
and in the CMB literature~\cite{pola,huwhite97}, specially because we
want to present both in a unified framework and language. From now on,
we use the second decomposition and its interest will be enlightened
by the study of the CMB anisotropies.

\subsection{CMB temperature and polarisation anisotropies}
\label{par3_2}

Gravitational waves, being metric perturbations, have an effect on the
temperature and polarisation of the CMB photons. Any metric perturbation
induces a fluctuation on the CMB temperature $\Theta$ through the
Sachs-Wolfe effect~\cite{sachs66} and any anisotropic distribution of
photons scattered by electrons will become polarised and
vice-versa. Since Thomson scattering generates linear polarisation,
we only need to consider the Stokes parameters $Q$ and $U$ and more
conveniently their two combinations $Q\pm iU$ which are invariant
under rotation.

Following Hu and White~\cite{huwhite97}, we decompose the tensorial
part of the temperature anisotropies according to
\begin{eqnarray}\label{dec_theta}
\Theta(\eta,{\bf x},\hat n) & = & \int \frac{\ddd^3{\bf k}}{(2\pi)^3}
\sum_\ell \sum_{m=\pm 2} \Theta_\ell^{(m)}(k,\eta) 
G_\ell^m({\bf k},{\bf x},\hat n), \\
\label{dec_qu}
(Q \pm i U)(\eta, {\bf x} , \hat n) & = & \int \frac{\ddd^3{\bf
k}}{(2\pi)^3}
\sum_\ell \sum_{m=\pm 2} (E_\ell^{(m)} \pm i B_\ell^{(m)}) (k,\eta)
\;_{\pm 2}G_\ell^m({\bf k},{\bf x},\hat n).
\end{eqnarray}
The coefficients $E_\ell^{(m)}$ and $B_\ell^{(m)}$ transform as $E_\ell
\to (-)^\ell E_\ell$ and $B_\ell \to - (-)^\ell B_\ell$ under parity and
are called the ``electric'' and ``magnetic'' part of the
polarisation. The functions $G_\ell^m$, $_{\pm 2}G_\ell^m$ form three
independent sets of orthonormal functions and depend both on the
position ${\bf x}$ and on the direction of propagation of the photons
$\hat n$ and are defined as
\begin{eqnarray}
G_\ell^m({\bf k},{\bf x},\hat n) &\equiv& (-)^\ell
          \sqrt{\frac{4\pi}{2\ell+1}} Y_\ell^m(\hat n)\exp(i {\bf k} . 
{\bf x}), \\
_{\pm 2}G_\ell^m({\bf k},{\bf x},\hat n) &\equiv& (-)^\ell 
          \sqrt{\frac{4\pi}{2\ell+1}} \;_{\pm 2}Y_\ell^m(\hat n) 
\exp(i{\bf k}.{\bf x}),
\end{eqnarray}
where the functions $Y_\ell^m(\hat n)$ are the standard spherical
harmonics and the functions $_{\pm 2}Y_\ell^m(\hat n)$ are the
spin-weighted spherical harmonics~\cite{Yslm1,Yslm2,Yslm3}. Note that
the decomposition on the basis $Q_{ij}^{(\pm2)}$ in the previous
section is enlightened by the fact that
$Q_{ij}^{(m)}n^in^j=G^{(m)}_2$~\cite{pola,huwhite97}.

The angular correlation function of these temperature/polarisation
anisotropies are observed on a 2-sphere around us and can be
decomposed in Legendre polynomials $P_\ell$ as
\begin{equation}\label{dTT}
\left<\frac{\delta U}{T}(\hat\gamma_1)
\frac{\delta V}{T}(\hat\gamma_2)\right>=\frac{1}{4\pi}
\sum_\ell (2\ell+1)C_\ell^{UV} P_\ell(\hat\gamma_1.\hat\gamma_2),
\end{equation}
where $U$, $V$ stand for $\Theta$, $E$, or $B$. Now the brackets stand
for an average on the sky, \IE{} on all pairs
$(\hat\gamma_1,\hat\gamma_2)$ such that $\hat\gamma_1.\hat\gamma_2=
\cos\theta_{12}$. Using the orthonormality properties of the
eigenfunctions $G$, equations~(\ref{dec_theta}-\ref{dec_qu}) can be
inverted to extract the angular power spectra $C_\ell^{UV}$ of the
temperature and polarisation anisotropies as
\begin{equation}
T_0^2 (2\ell+1)^2 C_\ell^{UV}(\eta_0) = \frac{2}{\pi} \int \frac{\ddd
k}{k}
\sum_{m=\pm 2} k^3 U_\ell^{(m)} (\eta_0, k) V_\ell^{(m)\;*} (\eta_0, k).
\end{equation}

The equations of evolution of the temperature and polarisation
multipoles $\Theta_\ell^{(m)}$, $E_\ell^{(m)}$ and $B_\ell^{(m)}$ can
be obtained by decomposing the Boltzmann equation satisfied by the
photon (or neutrino) distribution function on the eigenfunctions $G$
to get
\begin{eqnarray}\label{ev_theta}
\dot \Theta_\ell^{(m)} & = & k \left[
  \frac{_0 \kappa_\ell^m}{2\ell-1} \Theta_{\ell-1}^{(m)} 
- \frac{_0 \kappa_{\ell+1}^m}{2\ell+3} \Theta_{\ell+1}^{(m)} 
\right] - \dot \tau \Theta_\ell^{(m)} + \delta_{\ell,2} S^{(m)}, \\
\label{ev_e}
\dot E_\ell^{(m)} & = & k \left[
  \frac{_2 \kappa_\ell^m}{2\ell-1} E_{\ell-1}^{(m)} 
- \frac{2m}{\ell(\ell+1)} B_\ell^{(m)} 
- \frac{_2 \kappa_{\ell+1}^m}{2\ell+3} E_{\ell+1}^{(m)} 
\right]
 - \dot \tau \left(E_\ell^{(m)} + \delta_{\ell,2} \sqrt{6} P^{(m)}\right), \\
\label{ev_b}
\dot B_\ell^{(m)} & = & k \left[
  \frac{_2 \kappa_\ell^m}{2\ell-1} B_{\ell-1}^{(m)} 
+ \frac{2m}{\ell(\ell+1)} E_\ell^{(m)} 
- \frac{_2 \kappa_{\ell+1}^m}{2\ell+3} B_{\ell+1}^{(m)} 
\right] - \dot \tau B_\ell^{(m)} ,
\end{eqnarray}
where
\begin{eqnarray}
S^{(m)} &\equiv& \dot \tau P^{(m)} - \dot H^{(m)}, \\
P^{(m)} &\equiv& 
\frac{1}{10}\left(\Theta_2^{(m)} - \sqrt{6} E_2^{(m)} \right), \\
_s\kappa_\ell^m &\equiv& 
\ell
\sqrt{\left(1-\frac{m^2}{\ell^2}\right)\left(1-\frac{s^2}{\ell^2}\right)}.
\label{defkappa}
\end{eqnarray}
The differential optical depth $\dot \tau$ vanishes for neutrinos
(except in the very early universe, but the observable wavelengths in
the CMB are not affected by this) and is proportional to the free
electron density in the case of photons. It has to be calculated by
solving the relevant kinetic recombination equations for hydrogen and
helium~\cite{peebles,seager,bert}.

The quantity $P^{(m)}$ represents the coupling between temperature and
polarisation. Due to our choice of decomposition, only the electric
part of the polarisation is affected by temperature
anisotropies. However, electric and magnetic part of polarisation
couple themselves as photons propagate. The Clebsch-Gordan
coefficients $_s\kappa_\ell^m$ arise from product properties of
spherical harmonics. They are obtained in the same way as for the
scalar modes~\cite{bert}.
\begin{eqnarray}
\label{isw_tens}
\frac{\Theta_\ell^{(m)} (\eta_0, k)}{2\ell + 1} & = & 
\int_0^{\eta_0} \ddd\eta\; e^{-\tau}  S^{(m)} j_\ell^{(m)} (k(\eta_0-\eta)), \\
\label{pol_e}
\frac{E_\ell^{(m)} (\eta_0, k)}{2\ell + 1} & = & -\sqrt{6} 
\int_0^{\eta_0} \ddd\eta\; \dot\tau e^{-\tau} P^{(m)} 
\epsilon_\ell^{(m)} (k(\eta_0-\eta)), \\
\label{pol_b}
\frac{B_\ell^{(m)} (\eta_0, k)}{2\ell + 1} & = & -\sqrt{6} 
\int_0^{\eta_0} \ddd\eta\; \dot\tau e^{-\tau} P^{(m)} 
\beta_\ell^{(m)} (k(\eta_0-\eta)).
\end{eqnarray}
The functions $j_\ell^{(m)}$, $\epsilon_\ell^{(m)} $ and
$\beta_\ell^{(m)}$ are defined in terms of the spherical Bessel
functions, $j_\ell(x)$ as
\begin{eqnarray}
j_\ell^{(\pm 2)}(x) &\equiv&
 \sqrt{\frac{3}{8}\frac{(\ell+2)!}{(\ell-2)!}} \frac{j_\ell(x)}{x^2}, \\
\epsilon_\ell^{(\pm 2)}(x) &\equiv&
 \frac{1}{4} \left[-j_\ell(x) + j_\ell''(x)
+2\frac{j_\ell(x)}{x^2}+4\frac{j'_\ell(x)}{x}  \right], \\
\beta_\ell^{(\pm 2)}(x) &\equiv&
 \pm\frac{1}{2} \left[j_\ell'(x)+2\frac{j_\ell(x)}{x} \right]
\end{eqnarray}
where a prime denotes a derivative with respect to the argument $x$.

Beside the small contribution due to the polarisation in
\EQN{(\ref{isw_tens})}, the temperature fluctuation in the direction
$\hat\gamma$ reduce to the well known result by Sachs and
Wolfe~\cite{sachs66}
\begin{equation} 
\frac{\delta T}{T}(\hat\gamma) =
-\frac{1}{2}\int_{\eta_\LSS}^{\eta_0} \dot h_{ij}\gamma^i\gamma^j
\ddd\eta,
\end{equation}
where the subscript $\LSS$ stands for last scattering surface. The
``visibility function'' $\dot\tau e^{-\tau}$ appearing in equations
(\ref{pol_e}-\ref{pol_b}) takes a non zero value only at the time of
decoupling so that, contrarily to temperature anisotropies which are
constantly generated by gravitational interactions with photons,
polarisation is generated only at the last scattering surface.


\section{Damping of the gravitational waves}
\label{sec_damping}

We first study the effect of the damping of the gravitational waves
due to the anisotropic stress of the photons. Such a damping of the
amplitude of gravitational waves in various viscous cosmic media has
been already discussed in~\cite{dimitro}; we give a description of
this damping in the formalism we use here in order to quantify
precisely its effect on CMB anisotropies. On subhorizon scales larger
than the diffusion length $\lambda_{\rm D}\equiv\dot\tau^{-1}$ of the
photons,
\IE{} such that
\begin{equation}\label{scales}
\Hconf_\EFF\equiv\tau_\Hconf^{-1}\ll k\ll \dot\tau,
\end{equation}
the set of equations~(\ref{ev_theta}-\ref{defkappa}) implies that
\begin{equation}\label{theta2}
\Theta_2^{(\pm2)}=-\frac{4}{3}\frac{\dot H}{\dot\tau}\quad
\hbox{and}\quad
E_2^{(\pm2)}=-\frac{\sqrt{6}}{4}\Theta_2^{(\pm2)}.
\end{equation}
Since $\pi^{(\pm2)}$ is proportional to $\Theta_2^{(\pm2)}$, we can
insert the former expressions in the gravitational waves evolution
equation~(\ref{evoH}) to get the back reaction of the anisotropic
stress
\begin{equation}
\ddot H+2\Hconf_\EFF\dot H+k^2H=-\frac{32}{15}\kappa Pa^2\frac{\dot
H}{\dot\tau}.
\end{equation}
Setting
\begin{equation}\label{defh2}
H\equiv \bar H h ,
\end{equation}
with $\bar H$ solution of the homogeneous equation of evolution (\IE{}
with $\pi^{(m)}=0$), we get
\begin{equation}\label{evol_h}
\bar H\ddot h+\left(2\dot{\bar H}+2\Hconf_\EFF\bar H+
\frac{32}{15}\kappa Pa^2\frac{\dot{\bar H}}{\dot\tau}\right)\dot h
+\frac{32}{15}\kappa Pa^2\frac{\dot{\bar H}}{\dot\tau}h=0.
\end{equation}
Now, (i) since $\dot{\bar H}\simeq k\bar H$, we deduce that $\dot{\bar
H}\gg\Hconf_\EFF\bar H$, (ii) since $\kappa Pa^2\simeq \Hconf_\EFF$,
it follows that $(32/15)\kappa Pa^2\dot{\bar H}/\dot\tau\simeq
\Hconf_\EFF^2/\dot\tau\ll k\bar H\simeq \dot{\bar H}$ and (iii) since
$\ddot h\simeq \dot h/\tau_{\rm D}$, $\bar H\ddot h\ll\dot{\bar H}\dot
h$ and in conclusion in the limit~(\ref{scales}) the equation of
evolution of the gravitational waves in presence of the anisotropic
stress~(\ref{evol_h}) reduces to
\begin{equation}
\dot h=-\frac{16}{15}{\kappa Pa^2}{\dot\tau}h.
\end{equation}
We deduce that a mode $k$ is damped from the time it enters the Hubble
radius, \IE{} $\eta\simeq k^{-1}$ since it happens during the
radiation era, to roughly the time when the anisotropic stress becomes
negligible, \IE{} approximately at the time of last scattering
$\eta_\LSS$. It follows that
\begin{equation}
\label{eq_damp}
h(k,\eta_\LSS)\simeq  \hbox{exp}\left(-\frac{16}{15}
\int_{1/k}^{\eta_\LSS}\frac{\kappa P a^2}{\dot\tau}d\eta\right)
h\left(k,1/k\right).
\end{equation}
This damping of the gravitational waves by the anisotropic stress of
the photon fluid is analogous to the damping of the scalar modes
(density fluctuations) known as the Silk damping~\cite{silk68} a
description of which, in the formalism used here, can be found
in~\cite{huwhite97}. Note however that the origin of the damping is
different.

\begin{figure}
\centerline{
\psfig{figure=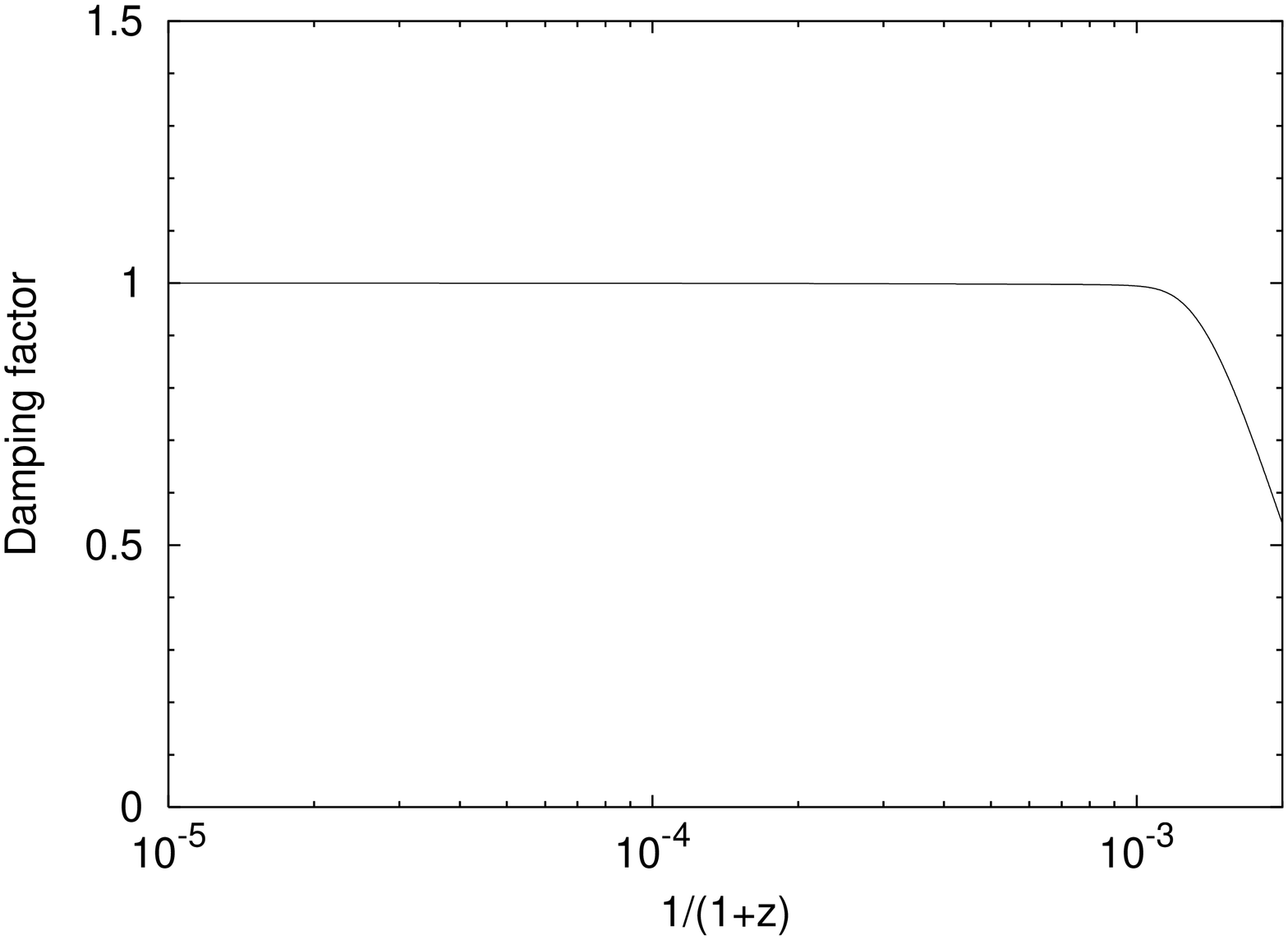,width=3.5in,angle=270}
\psfig{figure=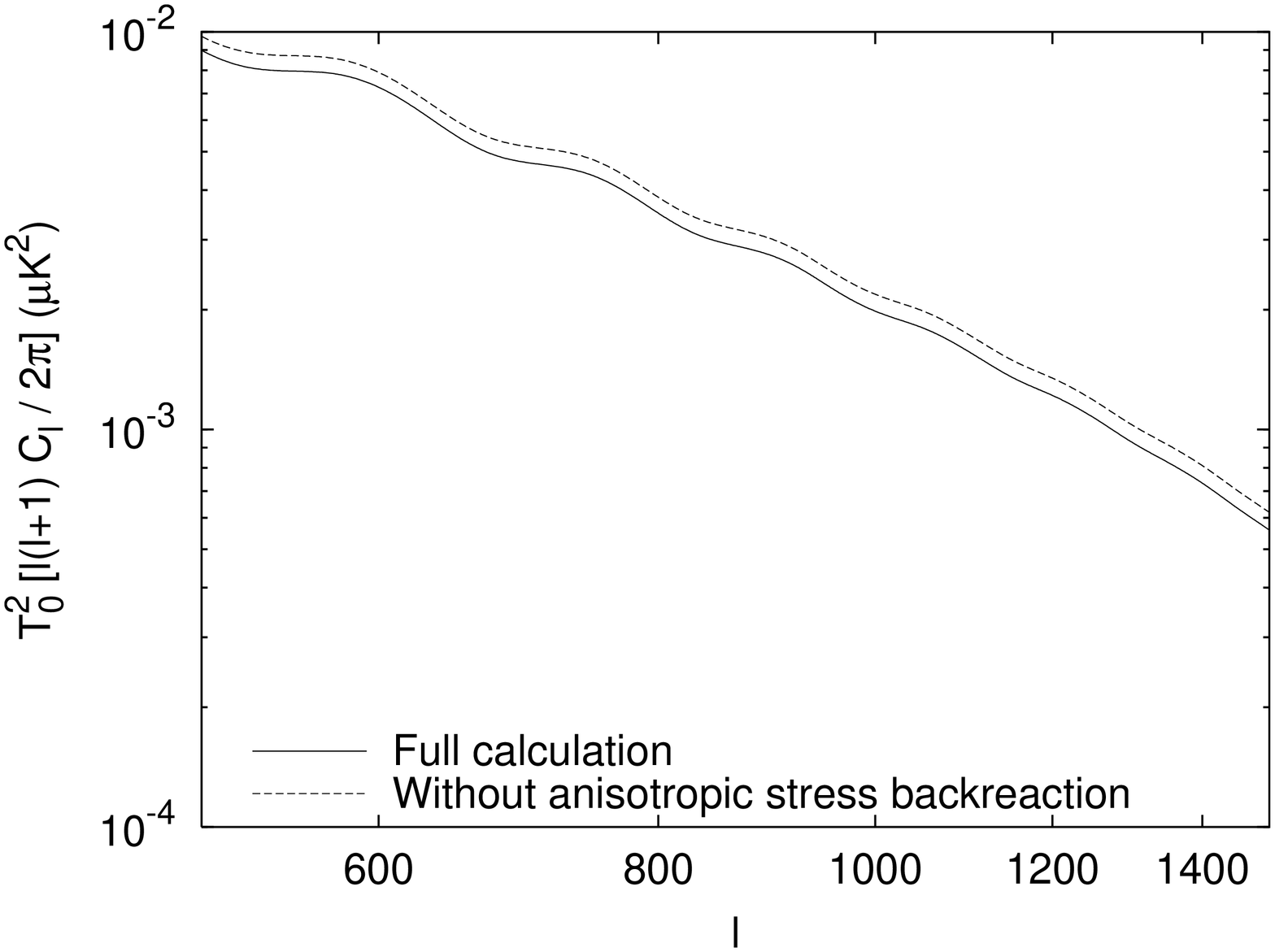,width=3.5in,angle=270}}
\caption{Damping of the gravitational waves due to its coupling to the
photons anisotropic stress. The damping (left figure) is generated
only when the universe becomes neutral, \IE{} soon before the last
scattering surface. As a consequence, all the mode which have already
entered into the Hubble radius at recombination are equally damped,
regardless of their wavelength. The influence of this damping on the
CMB anisotropies is shown on the right figure. Since all the modes are
equally damped, this translates into a constant ratio of amplitude
between the damped and the undamped cases. Note however that our
derivation is valid only when \EQN{(\ref{scales})} applies, which is
not true at the end of decouling, when $\dot\tau$ becomes small. This
is why the actual damping ($10\%$) is smaller than what is expected
from the left plot.}
\label{fig_damp}
\end{figure}

This effect is small but, apart from~\cite{dimitro}, was not much
emphasized in the literature before. Assuming that the universe is
completely ionized until the last scattering surface, the integral of
\EQN{(\ref{eq_damp})} is of order~\cite{huthesis}
\begin{equation}
\label{damp_fac}
\frac{1}{3}
\left(1-\frac{Y_{\rm He}}{2}\right)
\frac{\Omega_\gamma^0}{\Omega_b^0}
\frac{m_{\rm p} \kappa}{\sigma_{\rm Th}} a_0(\eta_\LSS - 1/k) 
\simeq  10^{-3} (1-1/k\eta_\LSS),
\end{equation}
where $m_{\rm p}$ is the proton mass, $\sigma_{\rm Th}$ is Thomson
scattering cross section. The real damping factor is greater than the
estimate~(\ref{damp_fac}) because the universe becomes neutral at the
last scattering surface (so that the term $\dot \tau$ is smaller). In
\FIG{\ref{fig_damp}}~[left], we plot this damping factor for the modes
that entered into the Hubble radius long before the last scattering
surface (\IE{} such that $k \gg \eta_\LSS^{-1}$). As a consequence, the
comparison between the damped case to the undamped case, shozn on
\FIG{\ref{fig_damp}}~[right] does not show significant differences. The
amplitude of the high-$\ell$ tail of the CMB anisotropy spectrum is
lowered by roughly $10\%$ when one includes this effect. The same
occurs of course for polarisation. We emphasize that this result does
not depend on any particular model, and is not included in the most
recent (3.2) version of CMBFAST.

\section{Specification of the model}
\label{par4}

\subsection{Model parameters}

At this stage, the model we are discussing depends on five parameters:
\begin{enumerate}
\item $f(\phi)$ which is an arbitrary function of the scalar field 
$\phi$,
\item $\xi$ the coupling of the scalar field with the background 
spacetime,
\item $\alpha$ the slope of the potential~(\ref{pot}) or~(\ref{potsugra}),
\item $\Omega_\phi^0$ the energy density in the scalar field today,
\item $P_T(k)$ the spectrum of the gravitational waves.
\end{enumerate}

Indeed there exist some constraints on these functions and parameters
and we make the following assumptions and choices:
\begin{enumerate}

\item We assume that $f(\phi)=\phi^2/2$; this is the only choice for 
which the coupling constant $\xi$ is dimensionless. Moreover such a
choice can be seen as the lowest term in an expansion of $f$ in powers
of $\phi$. As shown in~\cite{uzan99} there exists tracking solutions
for the field $\phi$ evolving in the potential~(\ref{pot}) with such a
coupling.

\item If the scalar field $\phi$ is coupled to the spacetime metric,
this coupling must be weak enough so that it does not generate a
significant time variation of the constants of nature~\cite{caroll}.
Taking into account the bound on the variation $|\dot G/G|$ of the
Newton constant~\cite{gillies} and on the variation $|\dot
\alpha/\alpha|$ of the fine structure constant~\cite{damour96}, it was
shown~\cite{chiba99} that
\begin{equation}\label{69}
-10^{-2}\lesssim\xi\lesssim 10^{-2}-10^{-1}.
\end{equation}
This bound is however sensitive to the shape of the potential. On the
other hand the experimental constraints (from the Shapiro effect and
the light deflection in the Solar system) on the post-Newtonian
parameters~\cite{MTW,damour92} imply~\cite{chiba99}
\begin{equation}
\left|\xi\right|\lesssim
\frac{3.9\times10^{-2}}{\sqrt{\alpha(\alpha+2)}}.
\end{equation}
However, in this class of models one does not try to have a theory
converging towards general relativity at late time and the coupling
$\xi$ is constant which is the main reasons of the severe bounds on
its value. This can be improved by generalising this kind of models by
considering them in the framework of scalar-tensor
theories~\cite{bartolo,bertolami}.

\item In most models $\alpha$ is not constrained theoretically. If the
matter content of the universe today is dominated by the
matter-radiation fluid then the fact that the
observations~\cite{perlmutter99} favour $-1<\omega_\phi<-0.6$ gives a
bound on $\alpha$, which is indeeed not the case anymore if the scalar
field starts to dominate. In \FIG{\ref{lamb}}~[left], we compare this
analytic estimate and the numerical determination of the energy scale
$\Lambda$ as a function of the slope $\alpha$. We see that if $\alpha>4$
then $\Lambda$ is at least larger than 1~TeV (when
$\Omega_{\phi}^0=0.7$).

\item The density parameter $\Omega_\phi^0$ is not severely constrained
theoretically, but observations seem to indicate $\Omega_\phi^0 \simeq
0.7$. One has to check that if the scalar field was dominating the
matter content of the universe at some early stage then it has to be
subdominant at the time of nucleosynthesis (see
\EG{}~\cite{ferreira}). The choice of $\Omega_\phi^0$ fixes the value of
the energy scale $\Lambda$ in~(\ref{pot}) or~(\ref{potsugra}); this is
the {\it coincidence problem}. On \FIG{\ref{lamb}}~[right], we depict
the variation of the energy scale $\Lambda$ with $\Omega_{\phi}^0$ and
$\alpha$. It is not very sensitive to $\Omega_{\phi}^0$ as long as
$0.1<\Omega_{\phi}^0<0.9$. In fact, when the quintessence field starts
to dominate the matter content and if we have reached the attractor then
$\ddd^2V/\ddd\phi^2\propto H^2$ (see~\cite{scaling2}), and
$H^2\simeq V/M_\PL^2$ so that we can estimate that the variation of
$\Lambda$ with $\alpha$ follows
\begin{equation}
\label{estim_lambda}
\Lambda=\left(\rho_{\rm crit}M_{\rm
Pl}^\alpha\right)^{\frac{1}{4+\alpha}}.
\end{equation}
We conclude that
$$
\frac{\delta\Lambda}{\Lambda}\sim
\frac{1}{4+\alpha}\frac{\delta\Omega_\phi^0}{\Omega_\phi^0}
$$
and thus a prescision of 10\% on $\Omega_\phi^0$ requires to tune
$\Lambda$ at a 1\% level if \EG{} $\alpha=6$, which is a less drastic
tuning than the usual cosmological constant fine tuning problem.

\begin{figure} 
\centerline{
\psfig{figure=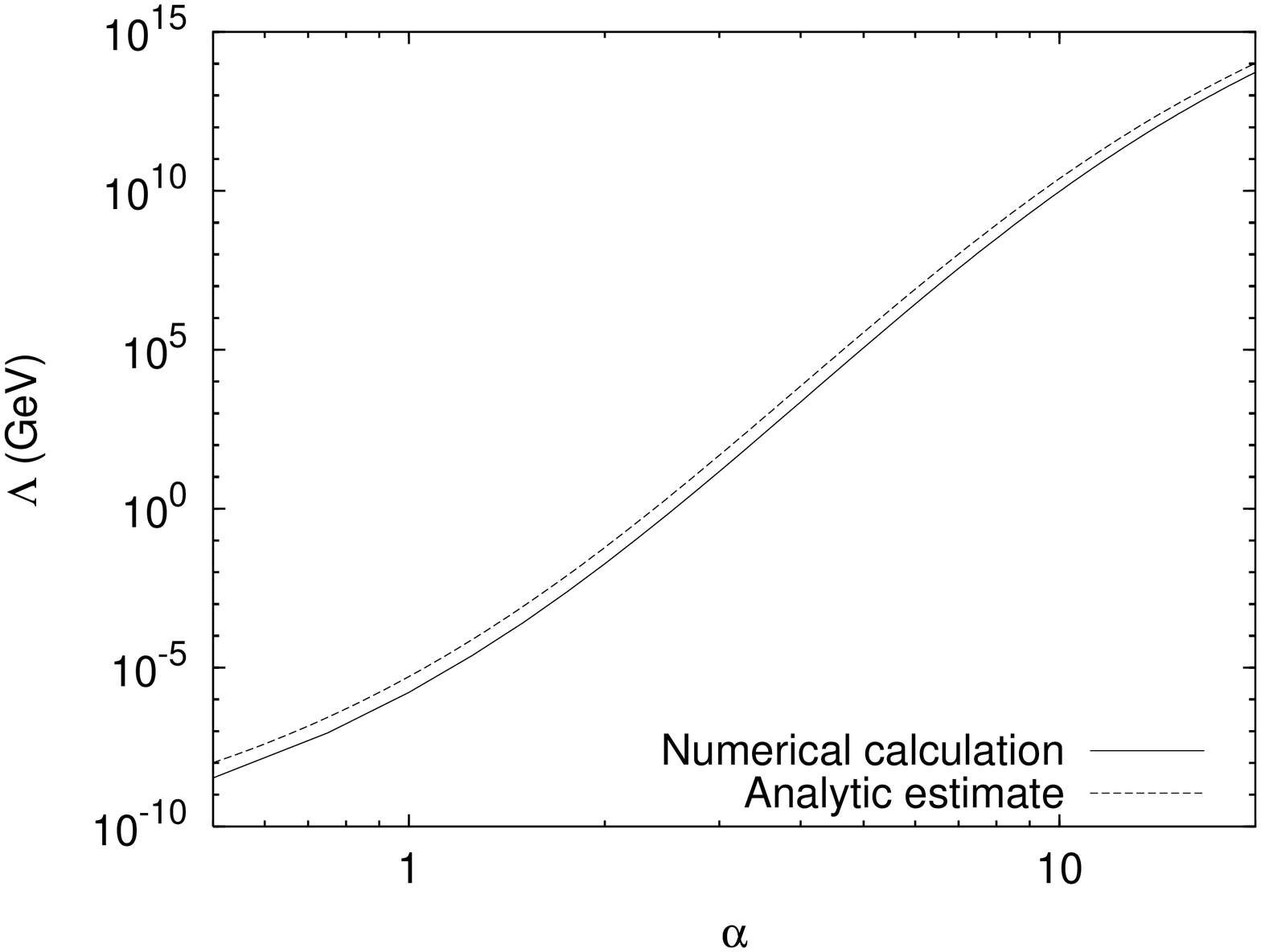,width=3.5in,angle=270}
\psfig{figure=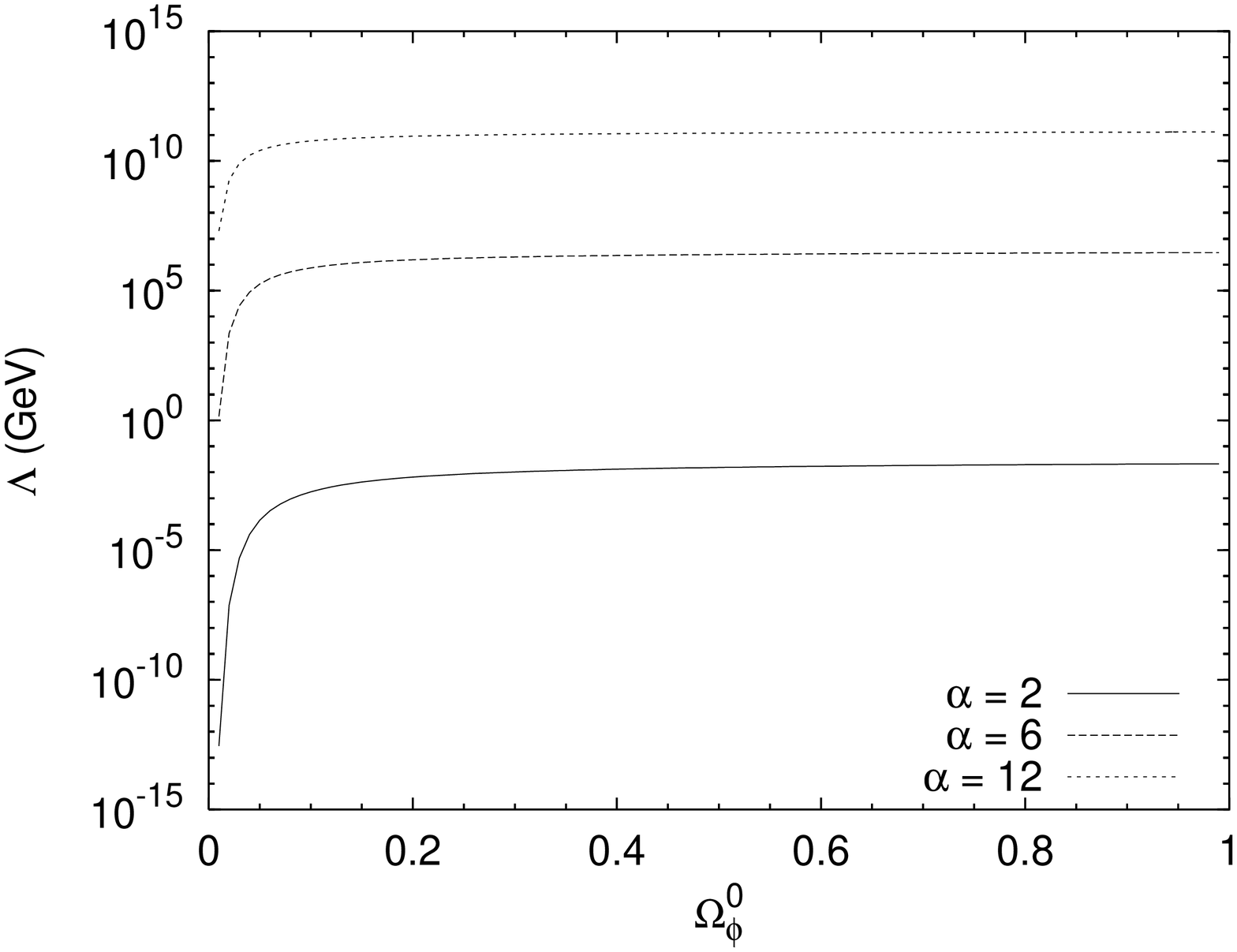,width=3.5in,angle=270}}
\caption{Variation of the energy scale $\Lambda$ of the potential
(\ref{pot}) with the energy density of the scalar field
$\Omega_{\phi}^0$ and the slope of the potential $\alpha$. We first
show [left] the variation of $\Lambda$ with $\alpha$ when
$\Omega_{\phi}^0=0.7$ and the comparison with the analytic
estimate~(\ref{estim_lambda}) and [right] the variation of $\Lambda$
with $\Omega_{\phi}^0$ for $\alpha=2,6,12$.}
\label{lamb} 
\end{figure} 

\item $P_T$ has to be determined by a specific model, such as
\EG{} inflation, and we parametrise it as
\begin{equation}
P_T(k)\equiv A_T k^{n_T}
\end{equation}
where $A_T$ is a constant dans $n_T$ is the tensor mode spectral
index. $A_T$ is obtained by normalising  the CMB temperature
anisotropies to the COBE data at $\ell=10$ for which
\begin{equation}
T_0 \sqrt{\frac{\ell(\ell+1)}{2\pi}C_\ell^{\Theta\Theta}}\simeq  30\,
\mu\hbox{K}.
\end{equation}
Since some measurements tend to show that there is a peak at the
degree scale~\cite{boomerang}, we conclude that a significant part of
the anisotropies may be generated by the scalar modes.  In the
``standard'' slow-roll inflation picture, this is compatible with an
almost scale-invariant spectrum with a low tensor contribution, in
which case the COBE results would put only an upper limit on the
amplitude of the gravitational waves spectrum. Nevertheless, we point
out that it is also possible that most of the large scale anisotropies
can be generated by gravitational waves. This assumes a strong
deviation from scale invariance ($n_S = 1.69$ and $n_T = 0.0$), but is
in good agreement with observational data~\cite{tegzeld}.
\end{enumerate}

\subsection{Initial conditions and behaviour of the background spacetime}

Concerning the initial conditions for the scalar field $\phi$, we will
consider the two extreme cases:
\begin{itemize}

\item \ICE{} where we assume that the scalar field is at equipartition
with the matter (\IE{} mainly with the radiation) deep in the
radiation era, 

\item \ICK{} where it dominates the matter content of the universe at
a very early stage.

\end{itemize}
Situation \ICE{} implies that at the end of reheating,
\begin{equation}
\rho_\phi\lesssim10^{-4}\rho_\gamma,
\end{equation}
where the factor $10^{-4}$ is roughly the inverse of the number of
degrees of freedom at that time. Since the quintessence field is
already subdominant at this epoch, one does not need to care about its
effect on nucleosynthesis since it remains subdominant until
recently. In the second situation \ICK, the field starts by dominating
and inflation ends by a kinetic phase rapidly than $\rho_\gamma$ and
will thus become subdominant. One has to check that this happens
before nucleosynthesis~\cite{giovannini99,vilenkin}. A realisation of
such initial conditions can be obtained in quintessential
inflation~\cite{vilenkin}.
\begin{figure} 
\centerline{
\psfig{figure=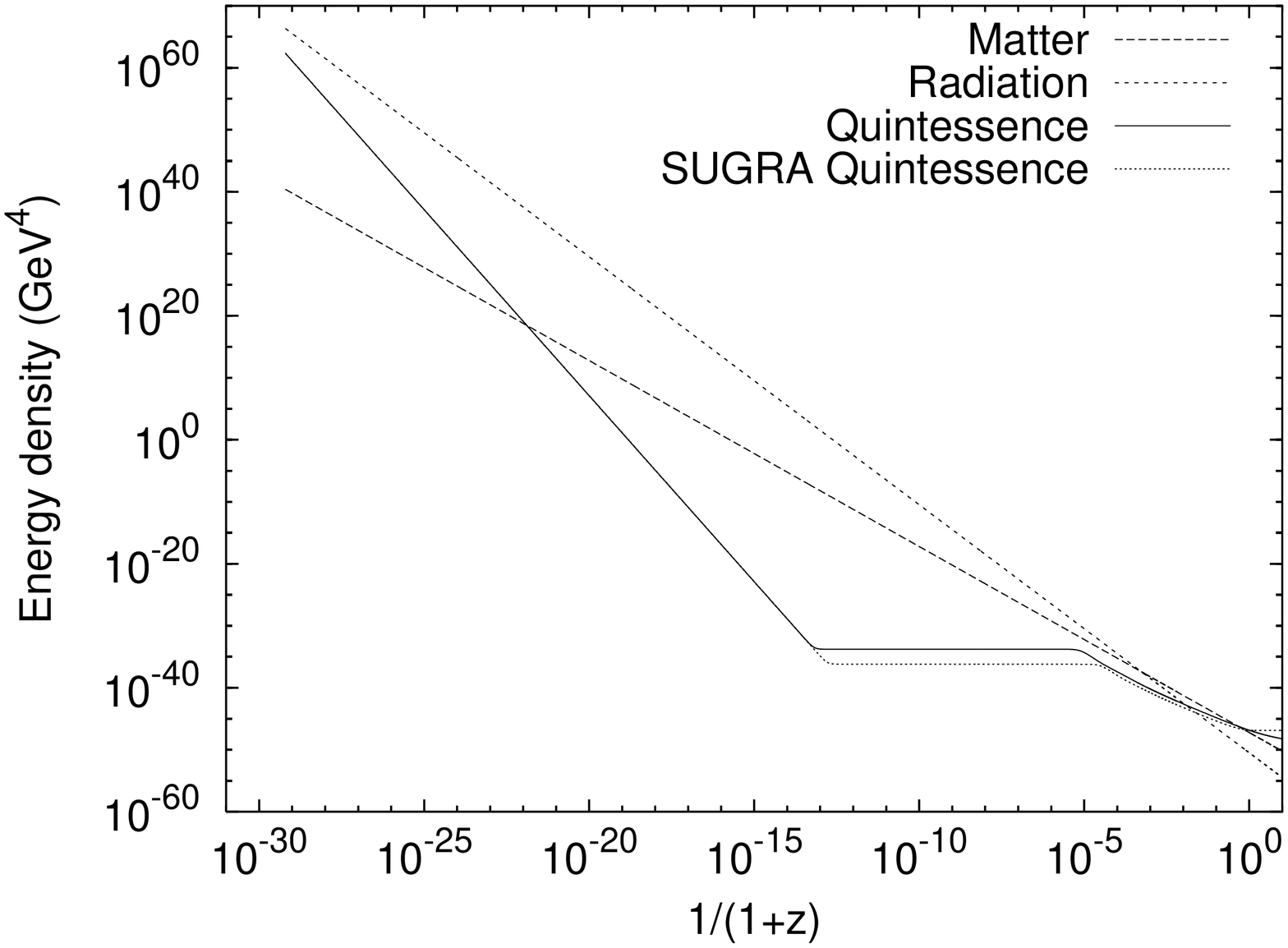,width=3.5in,angle=270}
\psfig{figure=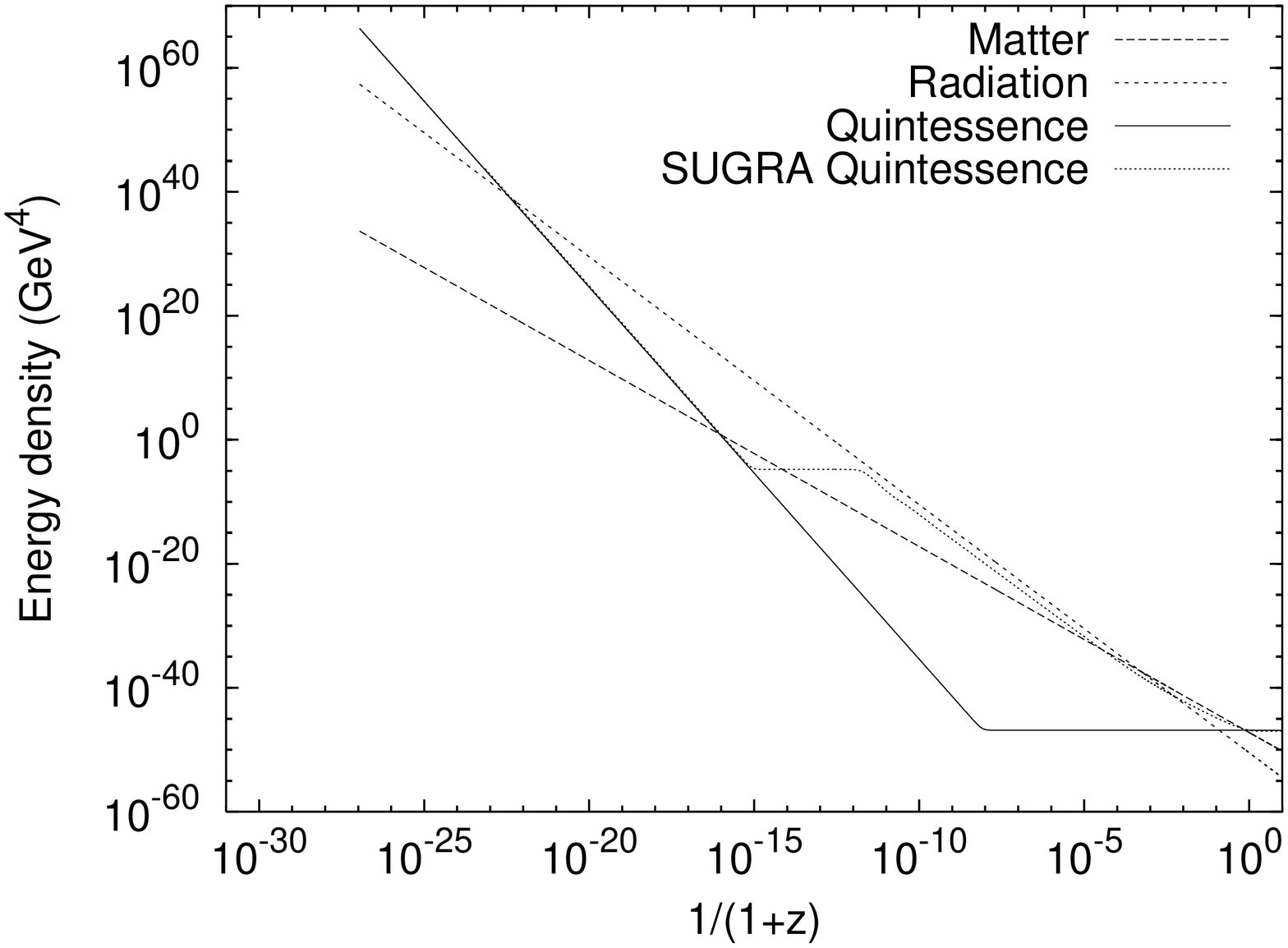,width=3.5in,angle=270}}
\caption{The behaviour of the energy density of the matter (long-dashed
line), radiation (short-dashed line) and scalar field (solid and dotted
lines), as a function of the redshift for the two class of initial
conditions: \ICE, when the field is at equipartition with the radiation
[left], and \ICK, when the field initially dominates the matter content
of the universe [right]. The solid line represents the case when the
field evolves in an inverse power-law potential, and the dotted line
represents the case when the field evolves in the SUGRA potential. Note
that when the field dominates at early times, the SUGRA potential
stabilizes the field, which reaches the tracking solution earlier.}
\label{plot1} 
\end{figure} 

\begin{figure} 
\centerline{
\psfig{figure=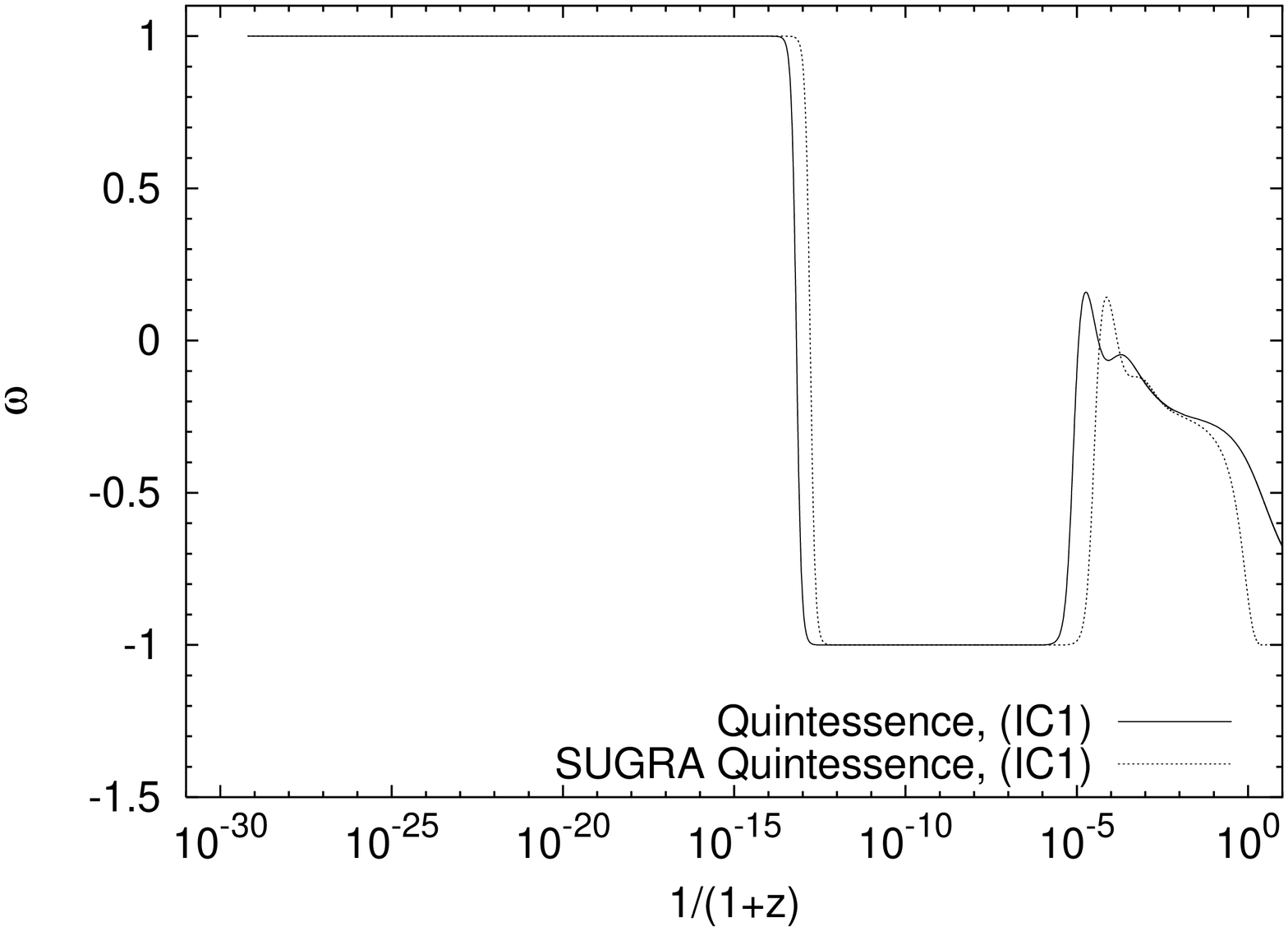,width=3.5in,angle=270}
\psfig{figure=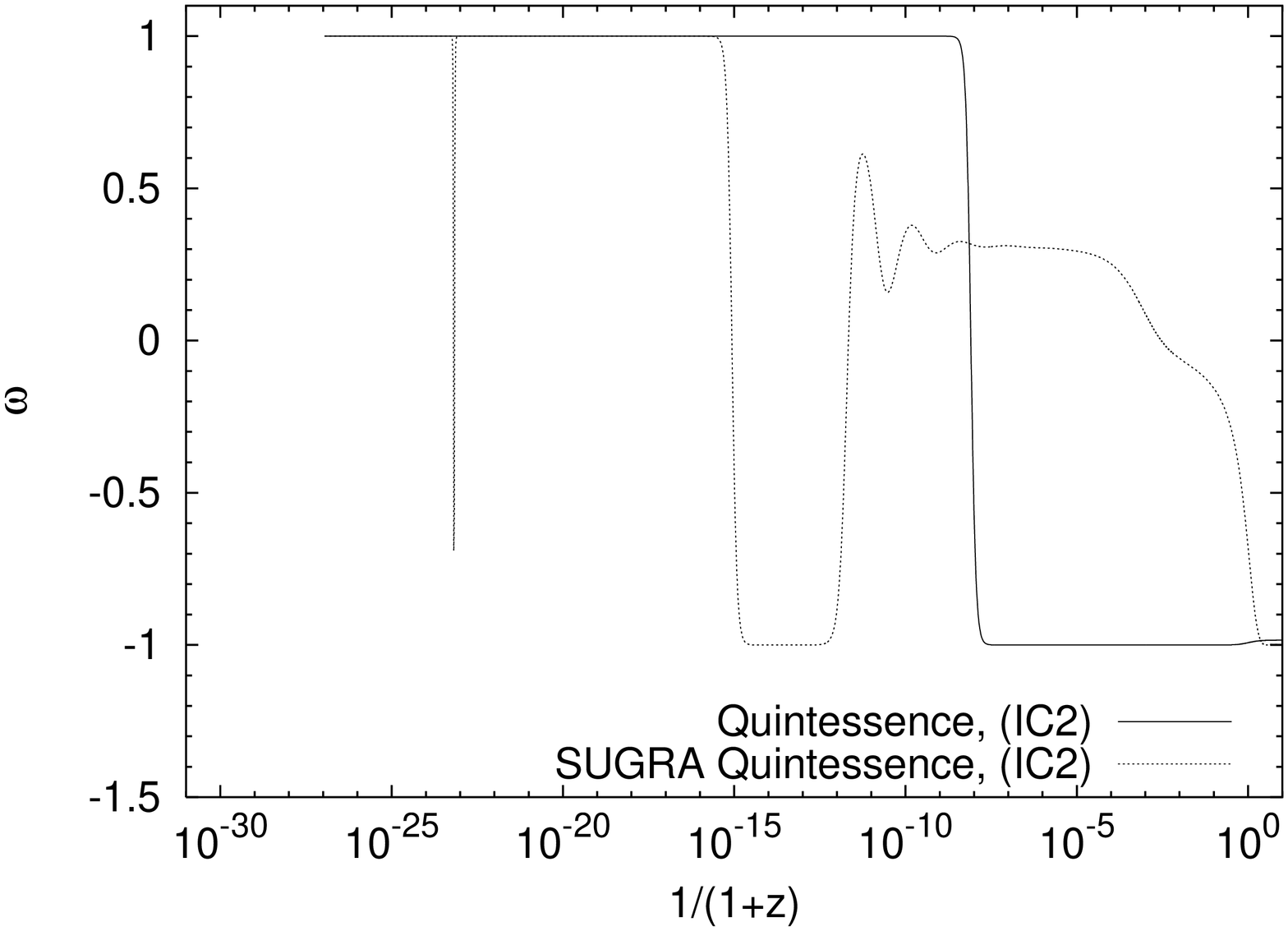,width=3.5in,angle=270}}
\caption{The behaviour of the equation of state parameter as a function
of the redshift for the two class of initial conditions: \ICE, when the
field is at equipartition with the radiation [left], and \ICK, when the
field initially dominates the matter content of the universe
[right]. The solid line represents the case when the field evolves in an
inverse power-law potential, and the dotted line represents the case
when the field evolves in the SUGRA potential. In the case of \ICK, the
field reaches the tracking solution only when SUGRA corrections to the
potential are considered. Note also the spikes in the SUGRA case
[right], which illustrate the fact that the field bounces around the
Planck scale (see \FIG{\ref{plot3}} below).}
\label{plot2} 
\end{figure} 

\begin{figure} 
\centerline{
\psfig{figure=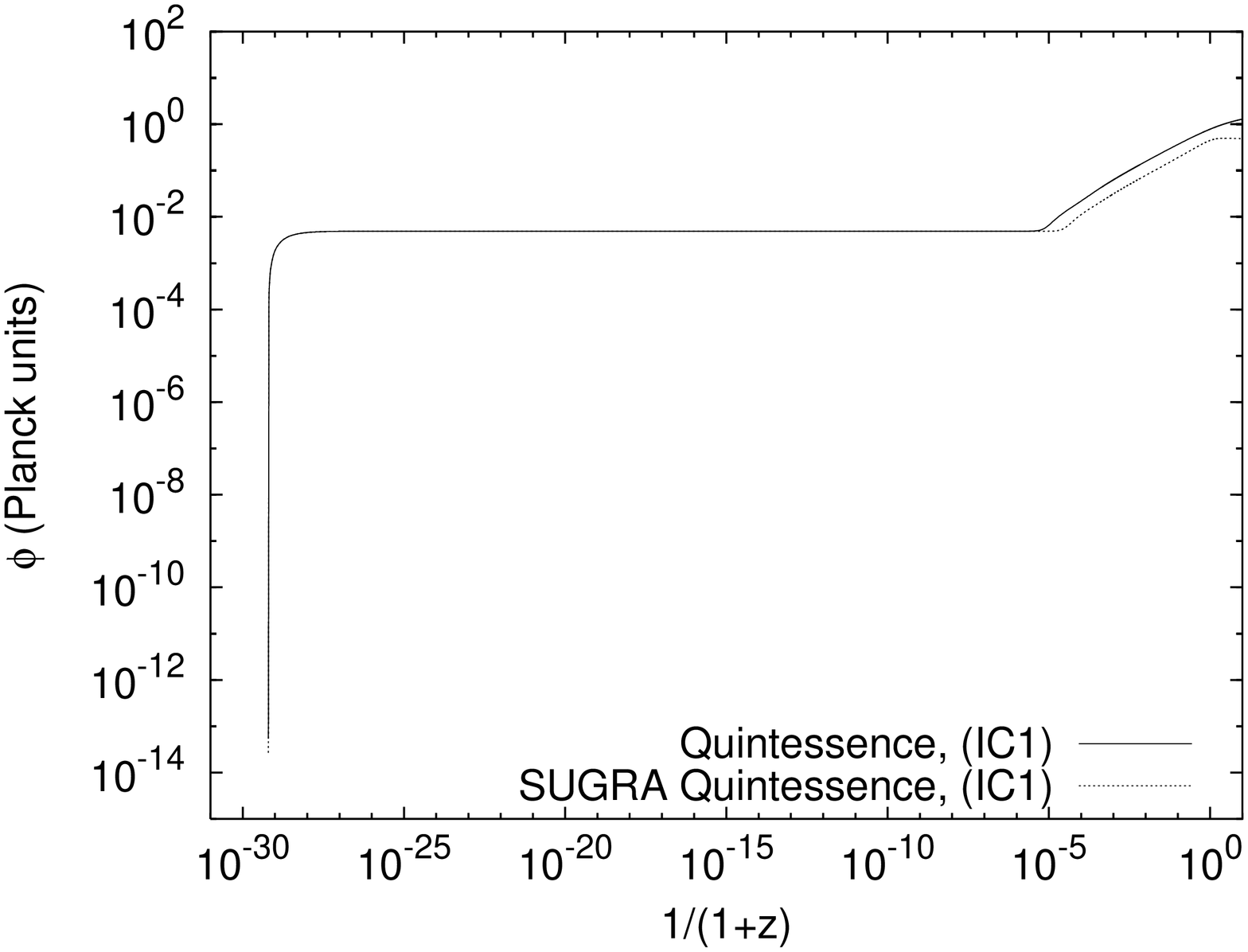,width=3.5in,angle=270}
\psfig{figure=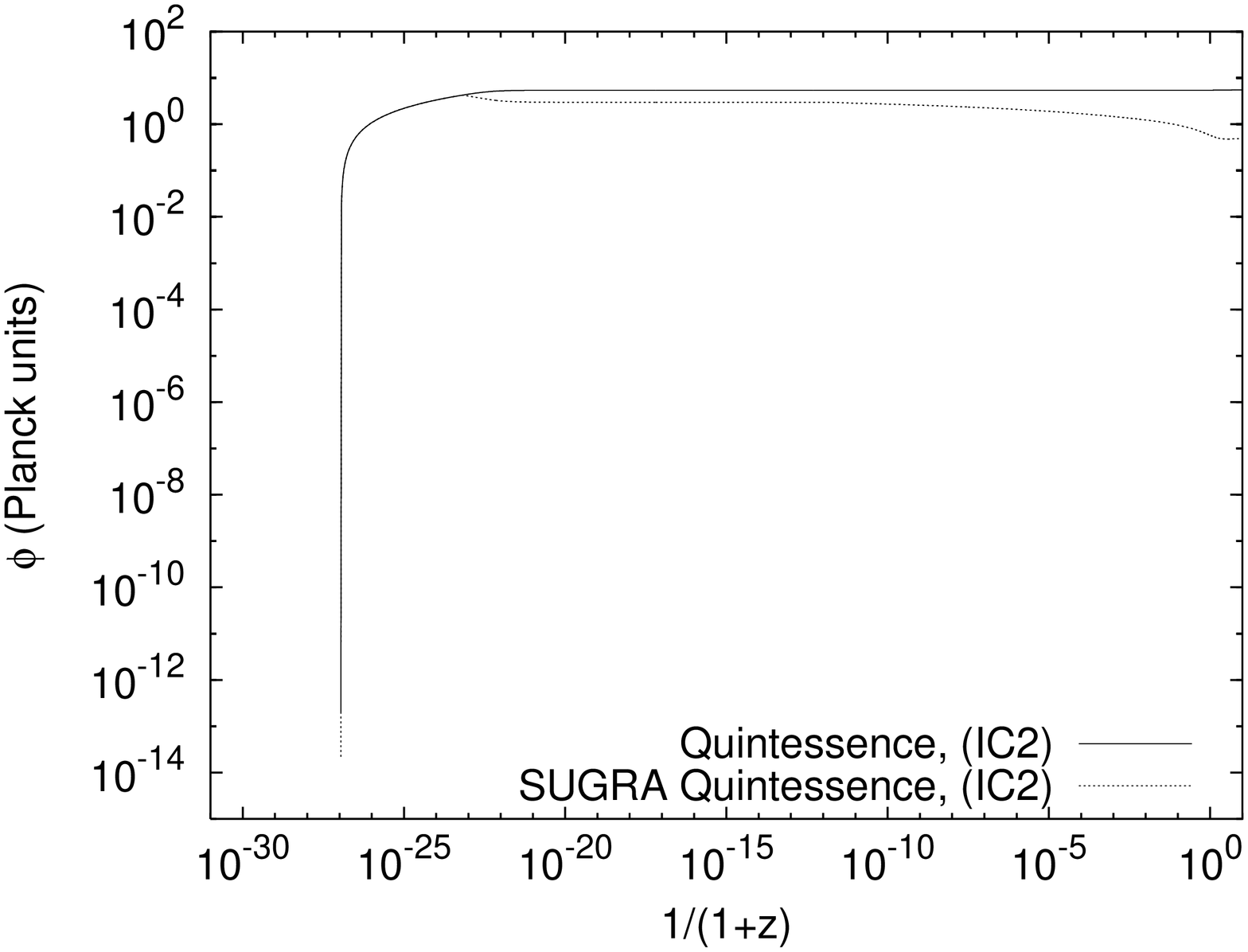,width=3.5in,angle=270}}
\caption{The behaviour of the quintessence field as a function of the
redshift for the two class of initial conditions: \ICE, when the field
is at equipartition with the radiation [left], and \ICK, when the field
initially dominates the matter content of the universe [right]. The
solid line represents the case when the field evolves in an inverse
power-law potential, and the dotted line represents the case when the
field evolves in the SUGRA potential. In the case of \ICE, the field
always reaches the tracking solution before today, whereas for \ICK, the
field reaches the tracking solution only when it evolves in the SUGRA
potential.}
\label{plot3} 
\end{figure} 

In \FIG{\ref{plot1}}, we depict the evolution of the energy density of
the quintessence field, matter and radiation for the initial conditions
\ICE~[left] and \ICK~[right]. We see that for a very large range of
initial conditions (roughly for $10^{-47}\hbox{GeV}^4\lesssim
\rho_\phi\lesssim10^{113}\hbox{GeV}^4$ at a redshift of $z\simeq
10^{30}$) we end up with a quintessence field which starts to dominate
today. This explains briefly how the fine tuning problem is
solved~\cite{zlatev}. We can also check that with these values the
scalar field does not dominate the matter content of the universe at
nucleosynthesis, \IE{} at a redshift of order $z\simeq 10^{10}$.

An interesting point concerns the evolution of the scalar field equation
of state in the case \ICK{} when $\xi=0$. The field rolls down very
fastly so that we are first in a regime where
\begin{equation}
\rho\simeq P\simeq\frac{1}{2}\frac{\dot\phi^2}{a^2}
\end{equation}
from which we conclude that its equation of state is
$\omega_\phi\simeq 1$ (see \FIG{\ref{plot2}}). But, due to the
exponential behaviour contribution of the potential, the field is
stopped when $\phi\gtrsim M_\PL$ and then rolls back to smaller values
(see \FIG{\ref{plot3}}) so that the field undergoes a series of damped
oscillations (because of the friction term coming from the expansion
in the Klein-Gordon equation). This implies that there exist times
such that $\dot\phi\simeq 0$ and thus small period around them where
the equation of state varies rapidly to $\omega_\phi\simeq -1$ (see
\FIG{\ref{plot2}}). This sudden change in the equation of state of
$\phi$ happen while it is dominating the matter content of the
universe (see \FIG{\ref{plot1}}) so that it implies variations in the
evolution of the scale factor of the universe which, in principle,
should let a signature in the gravitational waves energy
spectrum. Indeed, this does not happen in standard quintessence and is
a specific feature of the SUGRA-quintessence.

\vspace{0.5cm}

\begin{figure}
\centerline{
\psfig{figure=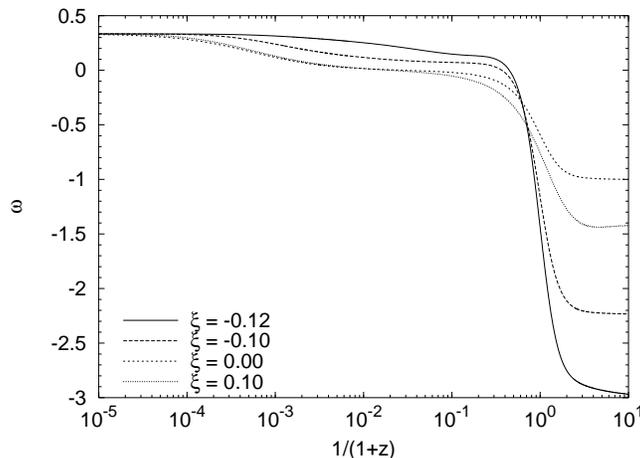,width=3.5in,angle=270}}
\caption{Evolution of the equation of state parameter $\omega$ as a
function of the redshift for different values of the coupling $\xi$. As
soon as the coupling is not minimal, $\omega$ can reach values smaller
than $-1$. The parameters of the model considered here are: $\alpha =
6$, potential~(\ref{potsugra}) including SUGRA corrections,
$\Omega_\phi^0 = 0.7$ and \ICE{} initial conditions.}
\label{fig_omega_xi}
\end{figure}

When $\xi \neq 0$, there are no significant modifications to the
background dynamics as long as the field has not reached the Planck mass
[because $2 \xi \kappa f(\phi)$ is small compared to unity, see
\EQN{\ref{kappa_eff}}]. Then, the main difference appears at late times
when the field starts to dominate and comes from the fact that the bound
$-1 < \omega_\phi < 1$ no longer applies, and one can get lower values
of $\omega_\phi$. Equivalently, the equation of state parameter $\omega
\equiv (P_\FLUID + P_\phi) / (\rho_\FLUID + \rho_\phi)$ for the whole
background fluids can reach values smaller than $-1$ (see
\FIG{\ref{fig_omega_xi}} where we plot the variation of $\omega_\phi$ as
a function of redshift). As pointed out by Caldwell~\cite{phantom}, such
a matter fits the current observational data. Different candidates such
as a decaying dark matter component~\cite{ziaeepour} and a kinetic
quintessence field~\cite{chiba2} were proposed. Here, we show that any
non-minimally coupled scalar field may be a good candidate for a
component of matter with $\omega<-1$. The constraints~(\ref{69}) on
$\xi$ implies that for our class of models
$$
-3\leq\omega_\phi<0
$$
if the scalar field dominates. We emphasize that $\omega_\phi$ is not
uniquely defined according to the way one splits $T^{\mu\nu}_{(\phi)}$
in~(\ref{Tphi}). In \FIG{\ref{fig_omega_xi}}, we used the Friedmann
equations (\ref{eq1}-\ref{eq2}) to extract $\omega$ from
$$
\frac{\dot{\cal
H}}{{\cal H}^2}-1=-\frac{3}{2}(1+\omega)\Omega
$$
and then $\omega_\phi$ from
$$
\omega\Omega=\sum_{i}
\omega_i\Omega_i
$$
where $i$ runs on all the matter species. This corresponds to the value
of $\omega$ as it may be reconstructed from observational data such as
\EG{} the supernovae type~Ia.

\section{Qualitative discussion}

\subsection{Gravitational waves spectrum}

Equation~(\ref{evoH}) describes the evolution of a damped oscillator.
Injecting the ansatz
\begin{equation}
H^{(m)}\equiv A^{(m)} \exp(ik\eta)
\end{equation}
in~(\ref{evoH}) and performing a WKB approximation leads to the
equation
\begin{eqnarray}\label{WKBs}
\dot A^{(m)}+\Hconf_\EFF A^{(m)} & = & 0
\end{eqnarray}
for the evolution of the amplitude $A^{(m)}$ where $\Hconf_\EFF\equiv
\dot{\widetilde a}/\widetilde a$, with
\begin{equation}
\widetilde a\equiv a\sqrt{1-2\kappa\xi f(\phi)}.
\end{equation}
This WKB approximation holds only for ``sub-horizon'' modes. Before a
mode has a wavelength smaller than the Hubble radius, its amplitude
evolves according to
\begin{equation}\label{WKBl}
\ddot A^{(m)}+2\Hconf_\EFF\dot A^{(m)}=0,
\end{equation}
the solutions of which are a constant mode and a decaying
mode. Neglecting the decaying mode, we see that the wave is ``frozen''
as long as its wavelength is larger than the Hubble radius, and that it
undergoes damped oscillations once its wavelength is shorter than the
Hubble radius. The damping of a mode of wavenumber $k$ between the time
it enters the Hubble radius and today is then proportional to
\begin{equation}
\frac{\widetilde a_k}{\widetilde a_0},
\end{equation}
where $\widetilde a_k$ is the scale factor evaluated at the time $\eta_k$
when the mode $k$ enters the Hubble radius (\IE{} when $\Hconf=k$) and
$\widetilde a_0$ is scale factor today. Injecting this behaviour in
(\ref{dOGW}), we obtain that the energy density spectrum of
gravitational waves scales as
\begin{equation}\label{scaling}
\frac{\ddd \Omega_\GW}{\ddd \ln(k)}\propto k^2 \widetilde a_k^2 P_T(k).
\end{equation}

First let us assume that $\xi=0$. For wavelengths corresponding to modes
that have entered the Hubble radius in the matter dominated era (for
which $a \propto\eta^2$ and thus $\eta_k\simeq k^{-1}$), one can easily
sort out that
\begin{equation}
\widetilde a_k\simeq  k^{-2}
\end{equation}
and the gravitational waves spectrum behaves as
\begin{equation}
\frac{\ddd \Omega_\GW}{\ddd \ln(k)} \propto k^{-2}  P_T(k).
\end{equation}
Equivalently, for wavelengths corresponding to modes entering the
Hubble radius in the radiation dominated era (for which $a \propto
\eta$) one can show that the gravitational waves energy spectrum
behaves as
\begin{equation}
\frac{\ddd \Omega_\GW}{\ddd \ln(k)} \propto k^0 P_T(k).
\end{equation}
To finish, if it happens that there exist wavelengths corresponding to
modes that have entered the Hubble radius while the scalar field was
dominating (for which $a \propto\sqrt{\eta}$ since $\rho_\phi \propto
1/a^6$) one obtains that
\begin{equation}\label{84}
\frac{\ddd \Omega_\GW}{\ddd \ln(k)} \propto k^1 P_T(k).
\end{equation}
In conclusion, we have found three behaviours for the gravitational
waves spectrum according to the wavelength. In \FIG{\ref{plot9}}, we
give an example of such a spectrum in a case where one has a scalar
field dominating at early stage [initial condition \ICK]. These
results hold also when $\xi\not=0$ but the slopes of the spectrum are
slightly changed since the time behaviour of $a$ has to be replaced by
the one of $\widetilde a$.

\subsection{CMB anisotropies}

For scales smaller than the Hubble radius at decoupling, one can follow
the same lines to predict the tensor part of the CMB temperature
anisotropies. The main difference is that the expression for
$\ell(\ell+1)C_\ell$ does not involve any factor $k^2$ as in
\EQNS{(\ref{dodlnk})}, the reason being that \EQN{(\ref{isw_tens})} can
be integrated by parts to drop the time derivative of $H^{(m)}$, which
shows that anisotropies are mostly generated on the last scattering
surface with an amplitude of $|H^{(m)}|^2$. Therefore, the spectrum
behaves as
\begin{equation}
\label{eq90}
\ell (\ell+1) C_\ell \propto \ell^{n_T-4},\ell^{n_T-2},\ell^{n_T-1},
\end{equation}
for modes which have entered the Hubble radius is the matter
dominated, radiation dominated and kinetic scalar field dominated eras
respectively. With standard cosmological parameters, the radiation to
matter transition occurs soon before the decoupling, and the scalar
field dominates only at very early times. As a consequence, one sees
almost only the regime $\ell (\ell+1) C_\ell \propto
\ell^{n_T-2}$. For modes which enter into the Hubble radius after the
last scattering surface, one can show~\cite{turner93} that the
produced spectrum scales as
\begin{equation}
\label{eq91}
\ell (\ell+1) C_\ell \propto \ell^{n_T}.
\end{equation}
Note that this expression is indeed an approximation and that it is
not easy to calculate an accurate analytical solution~\cite{grish93}.
These results are illustrated on \FIG{\ref{cdmplot}}. As already
stressed, the result of \EQN{\ref{eq91}} applies at large angular
scales which have not entered into the Hubble radius at
recombination. For standard cosmologies, this occurs for multipoles
smaller than $\ell \simeq 100$ (in addition, there are also some
corrections to this rough estimate which occur at the very smallest
multipoles and slightly boost the spectrum, as can also be seen on
\FIG{\ref{cdmplot}}). Then, at higher multipoles the result of
\EQN{\ref{eq90}} is valid. The matter dominated regime before
recombination is rather short, and occurs only between $\ell \simeq
100$ and $\ell \simeq 200$ (less than one oscillation in the
spectrum). For $\ell \gtrsim 200$, one sees the regime $\ell (\ell+1)
C_\ell \propto \ell^{n_T-2}$ (see also \FIG{1} of~\cite{turner93}).

\subsection{Results of the $\Lambda$CDM model}

\begin{figure} 
\centerline{
\psfig{figure=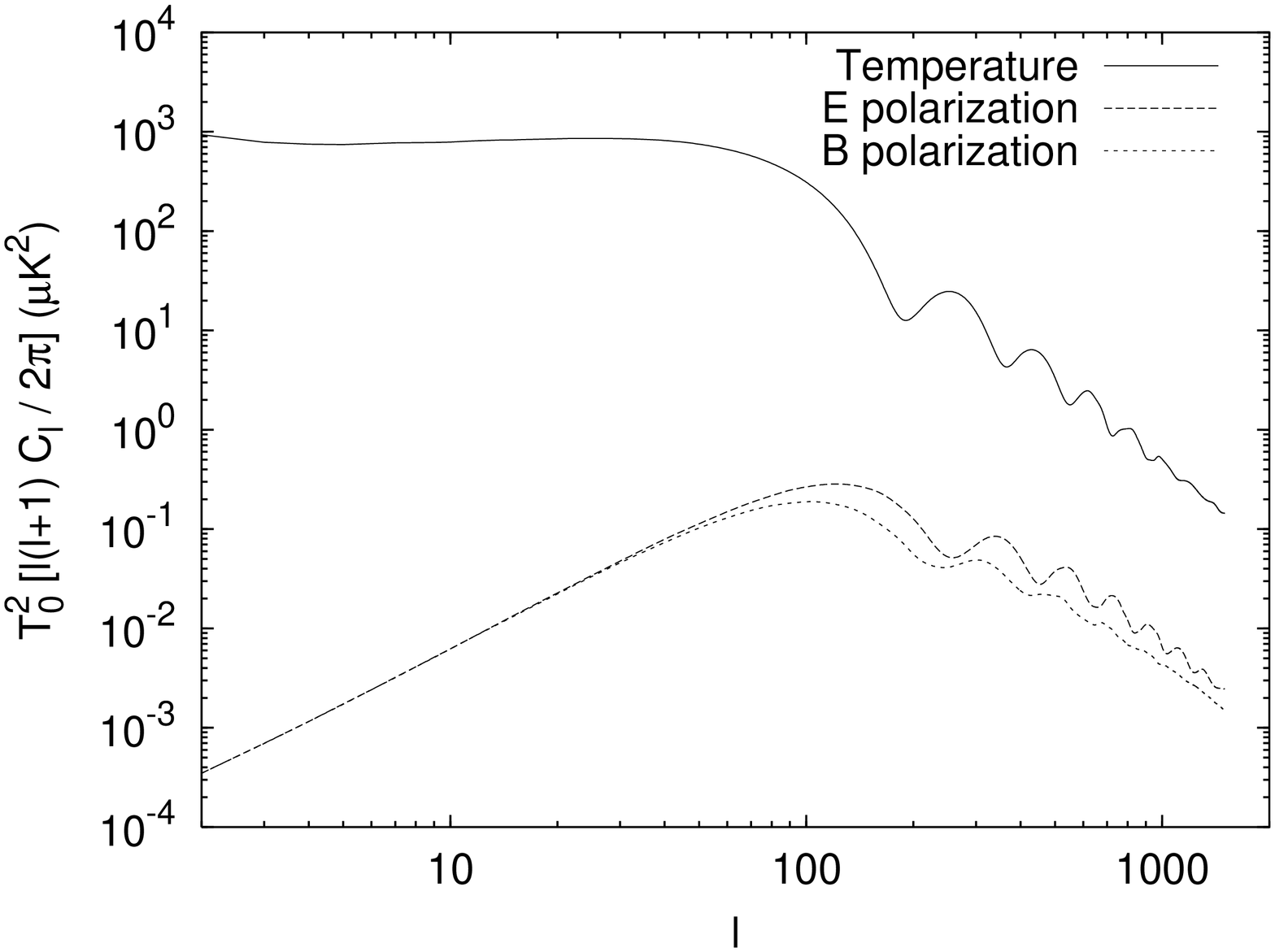,width=3.5in,angle=270}
\psfig{figure=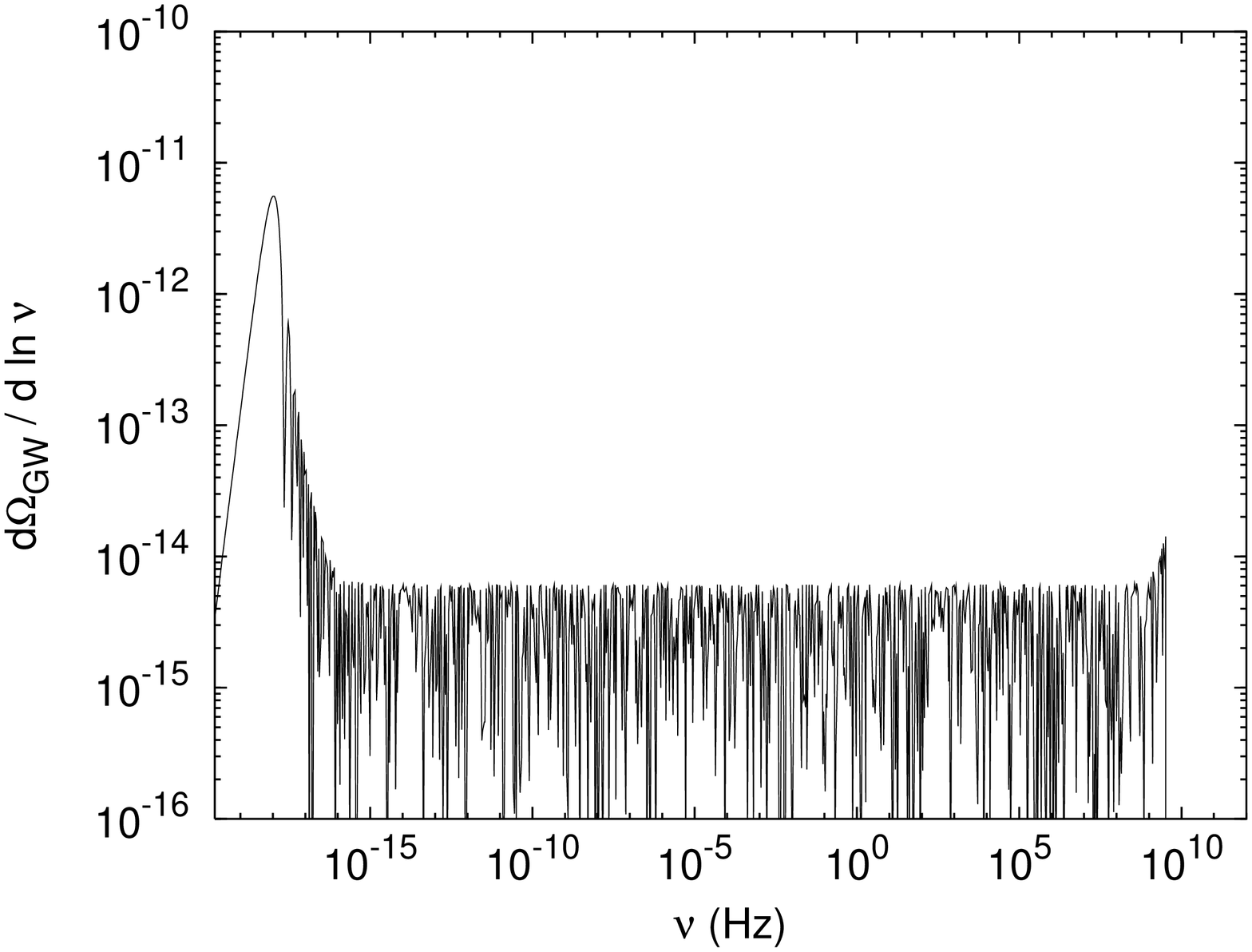,width=3.5in,angle=270}}
\caption{The temperature et polarisation of the CMB induced by
gravitational waves [left] and their energy density spectrum [right]
in a standard $\Lambda$CDM model with $n_T=0$.}
\label{cdmplot}
\end{figure}

Before turning to a more general numerical study of the class of
models we consider in this article, we recall in \FIG{\ref{cdmplot}}
the general results for the temperature and polarisation angular power
spectra and the gravitational waves density spectrum for a
$\Lambda$CDM model. This spectrum has two branches: a soft branch at
lower frequencies (corresponding to the matter dominated era) and a
hard branch at higher frequencies (corresponding to modes that entered
the horizon in the radiation era). Following~\cite{gw2}, we set the
cut-off on this spectrum to the last mode that has been inflated out
of the Hubble radius.

\section{Numerical results}
\label{par5}

\subsection{Field $\phi$ initially at equipartition}

Since the scalar field only starts to dominate at very recent time, we
expect no effect on the gravitational waves energy spectrum (since at
earlier time the universe is always radiation dominated). However, the
change in today's universe equation of state yield a specific
signature in the angular diameter-distance relation. Hence, one
expects to see the quintessence field behaviour in the positions of
the peaks in the CMB anisotropy spectra.

\begin{figure} 
\centerline{
\psfig{figure=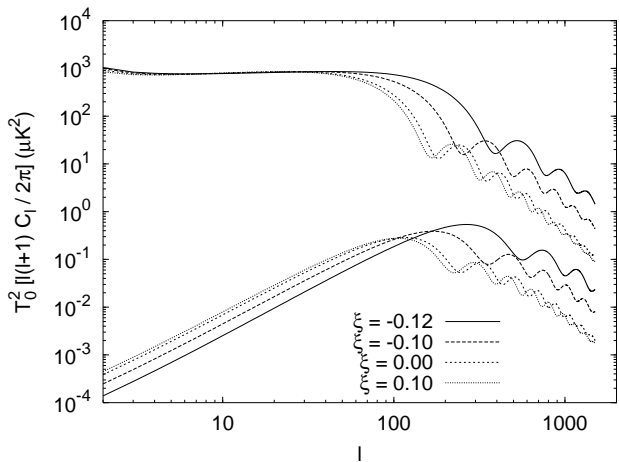,width=3.5in,angle=270}}
\caption{Influence of the coupling $\xi$ on the CMB temperature
and polarisation anisotropies. The value of $\xi$ influences the
angular diameter-distance relation and therefore affects the overall
position of the spectrum. The parameters of the model considered here
are the same as in \FIG{\ref{fig_omega_xi}}: $\alpha = 6$,
potential~(\ref{potsugra}) including SUGRA corrections, $\Omega_\phi^0
= 0.7$ and $n_T = 0$.}
\label{plot14} 
\end{figure}

The temperature anisotropies plots of \FIG{\ref{plot14}} are therefore
identical at high multipoles except for their overall position which
are different. At low redshift, the scalar field dominates and the
dynamics of the expansion depends explicitely of the value of the
coupling $\xi$, which cause some slight differences in the CMB
anisotropies at the very first multipoles ($\ell \leq 5$). We have
also seen that the polarisation is generated by gravity and therefore
different gravitational constants lead to different normalisation
between the polarisation and the temperature spectra. Since we
normalise the ``bare'' Einstein constant $\kappa$ so that the
effective Einstein constant corresponds to what we measure (in \EG{} a
Cavendish experiment), models with a different $\xi$ have different
$\kappa$. At decoupling, the scalar field does not dominate and
therefore $\kappa^\LSS = \kappa_\EFF^\LSS$. This induces different
amplitudes for the polarisation anisotropy spectra. For the lowest
values of $\xi$ there is a factor $2 - 4$ in amplitude as compared
with the $\xi = 0$ case, which roughly corresponds to the square of
the variation of $\kappa_\EFF$ (and, hence $G$) between the last
scattering surface and now. Note that the effect of $\xi$ depends on
its sign. This is the reason why the constraint derived by
Chiba~\cite{chiba99} are stronger for negative values of $\xi$. The
same can be seen in \FIG{\ref{fig_omega_xi}}.

We conclude that the temperature anisotropies and polarisation give
mainly information on the spectral index $n_T$, the energy density of
the scalar field today $\Omega_\phi^0$ and its coupling $\xi$.

\subsection{Field $\phi$ dominates at early stage}

We now turn to the more unusual case where the scalar field dominates
at the end of inflation and where the universe undergoes a kinetic
phase before the radiation era~\cite{ford87,spoko93} as in \EG{}
quintessential inflation~\cite{vilenkin}. The wavelengths
corresponding to the observable CMB multipoles ($\ell\lesssim2000$)
are much larger than the Hubble radius at nucleosynthesis, epoch at
which we have to be radiation dominated. As a consequence, we expect
no signature from this early phase on the CMB anisotropies and
polarisation.

\begin{figure}
\centerline{
\psfig{figure=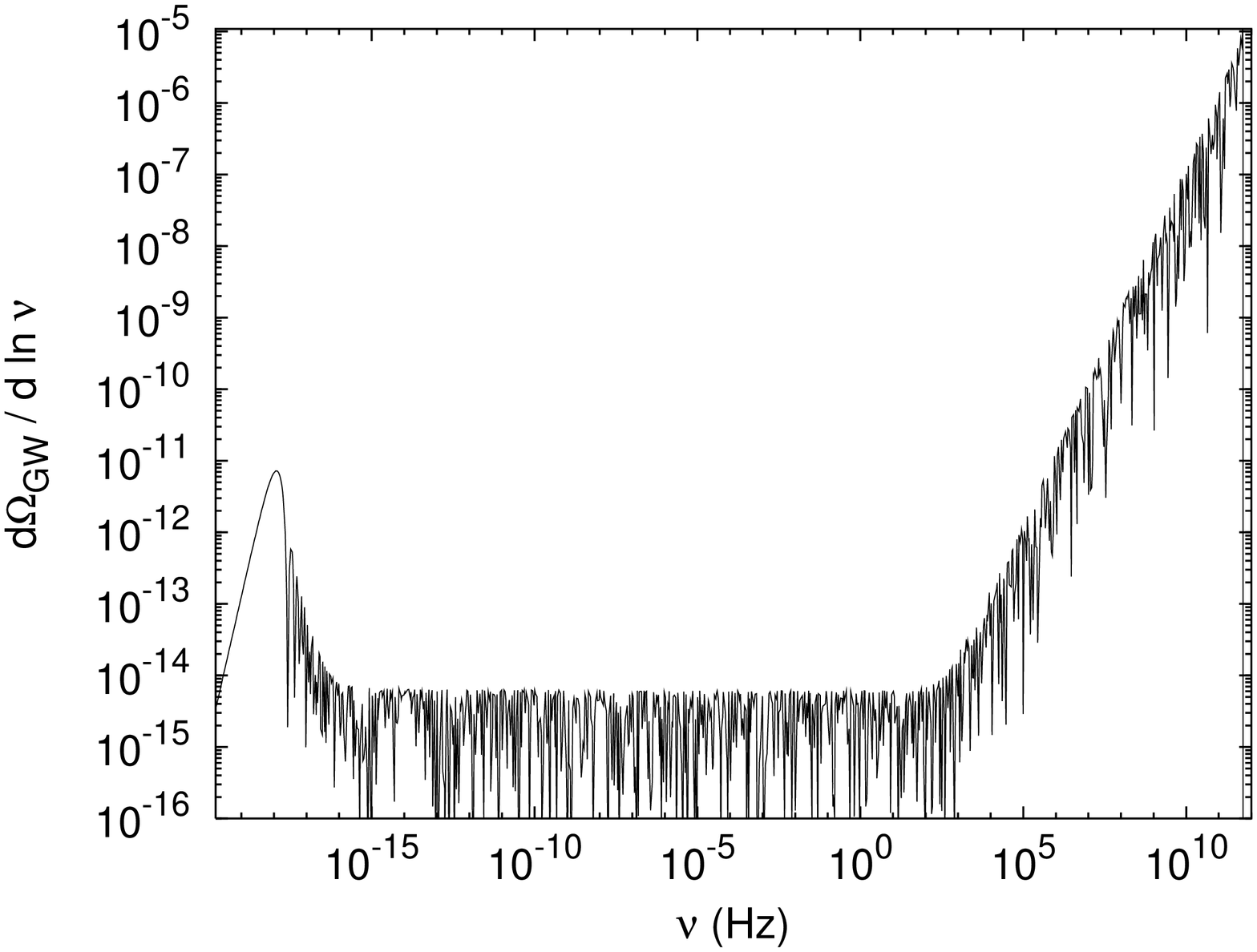,width=3.5in,angle=270}
\psfig{figure=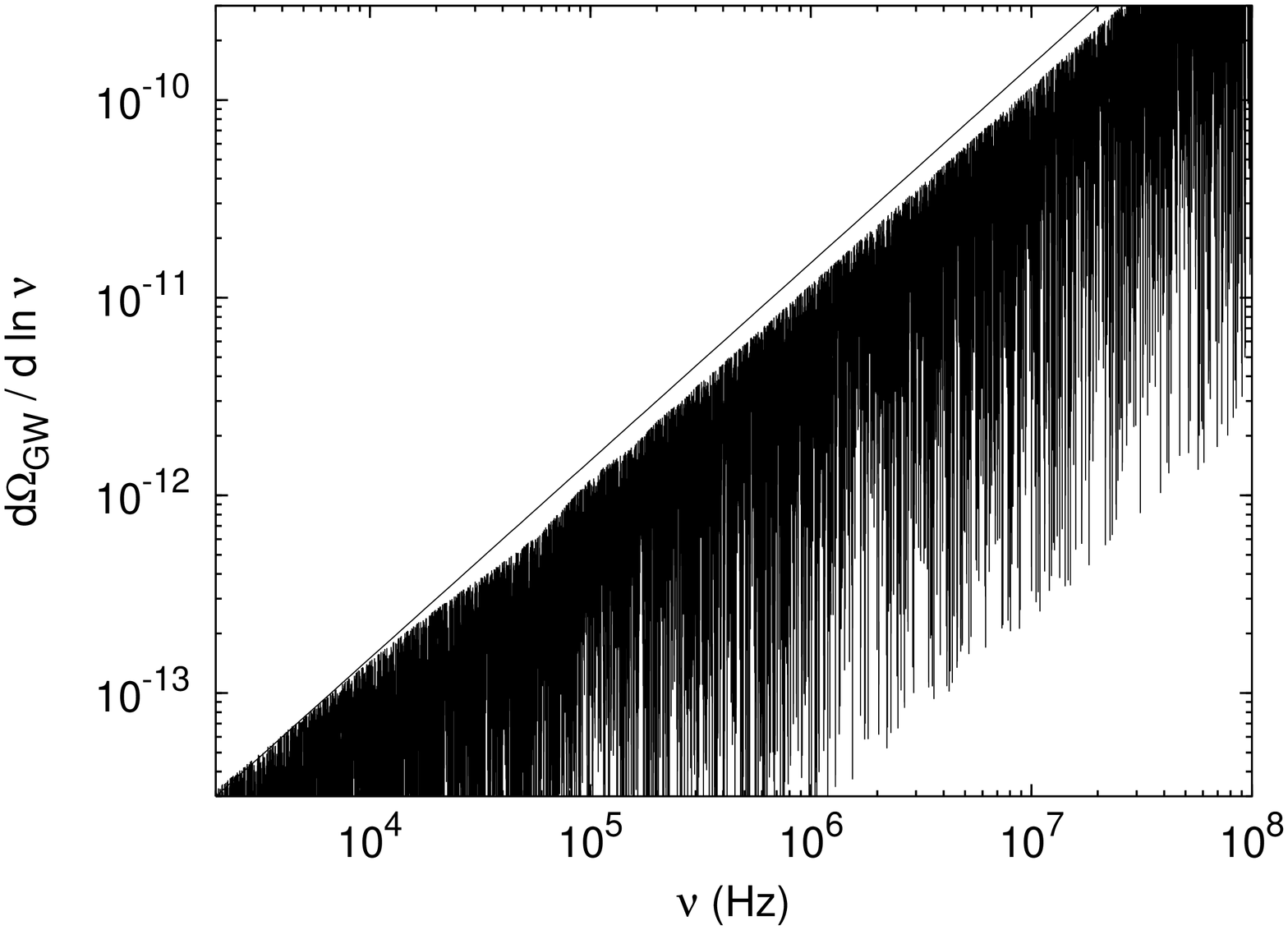,width=3.5in,angle=270}}
\caption{Gravitational waves spectrum in a quintessence model with
\ICK{} initial conditions with a SUGRA-like potential. The spectrum
has been normalized to be compatible with COBE at large scales. The
spike in the evolution of the equation of state of the scalar field
(see figure~\ref{plot2}) yields to a loss [right] of about 20\% in the
amplitude of the spectrum at high frequency (\IE{} $> 10^4\,{\rm
Hz}$).}
\label{plot9}
\end{figure}

As first pointed out in~\cite{giovannini99,vilenkin}, if the scalar
field dominates at early stage, there is an excess of gravitational
waves at high frequency [see equation~(\ref{84})]. On
\FIG{\ref{plot9}}, we present such a spectrum and we will discuss the
implication of this excess later.

An interesting effect concerns the difference between the spectra
obtained from an inverse power law potential and a SUGRA-like
potential. As shown on \FIG{\ref{plot9}}~[right], the amplitude of the
spectrum at high frequency in roughly 30\% higher for inverse power
law potentials. The relative decrease in amplitude at these
frequencies for SUGRA-like potentials depends on the dynamics of the
scalar field in the bounce (see \FIGS{\ref{plot2} and~\ref{plot3}})
during which the equation of state varies from $+1$ to $-1$ and to
$+1$ again. Thus, during this time, the modes that had just entered
into the Hubble radius (and thus which had just started to undergo
damped oscillations) went out of it (during the $\omega<0$ epoch) and
their amplitude was frozen before re-entering the Hubble radius
again. Hence, the modes of larger wavelengths are less damped which
explains this decrease in amplitude. Now, if the slope of the
potential is less steep, the bounce lasts longer (note that we always
reach $\omega=-1$ at the point where $\dot\phi=0$) and thus the
damping is stronger. This signature, even if not detectable by coming
experiment is nevertheless a clear feature of supergravity.

\begin{figure}
\centerline{
\psfig{figure=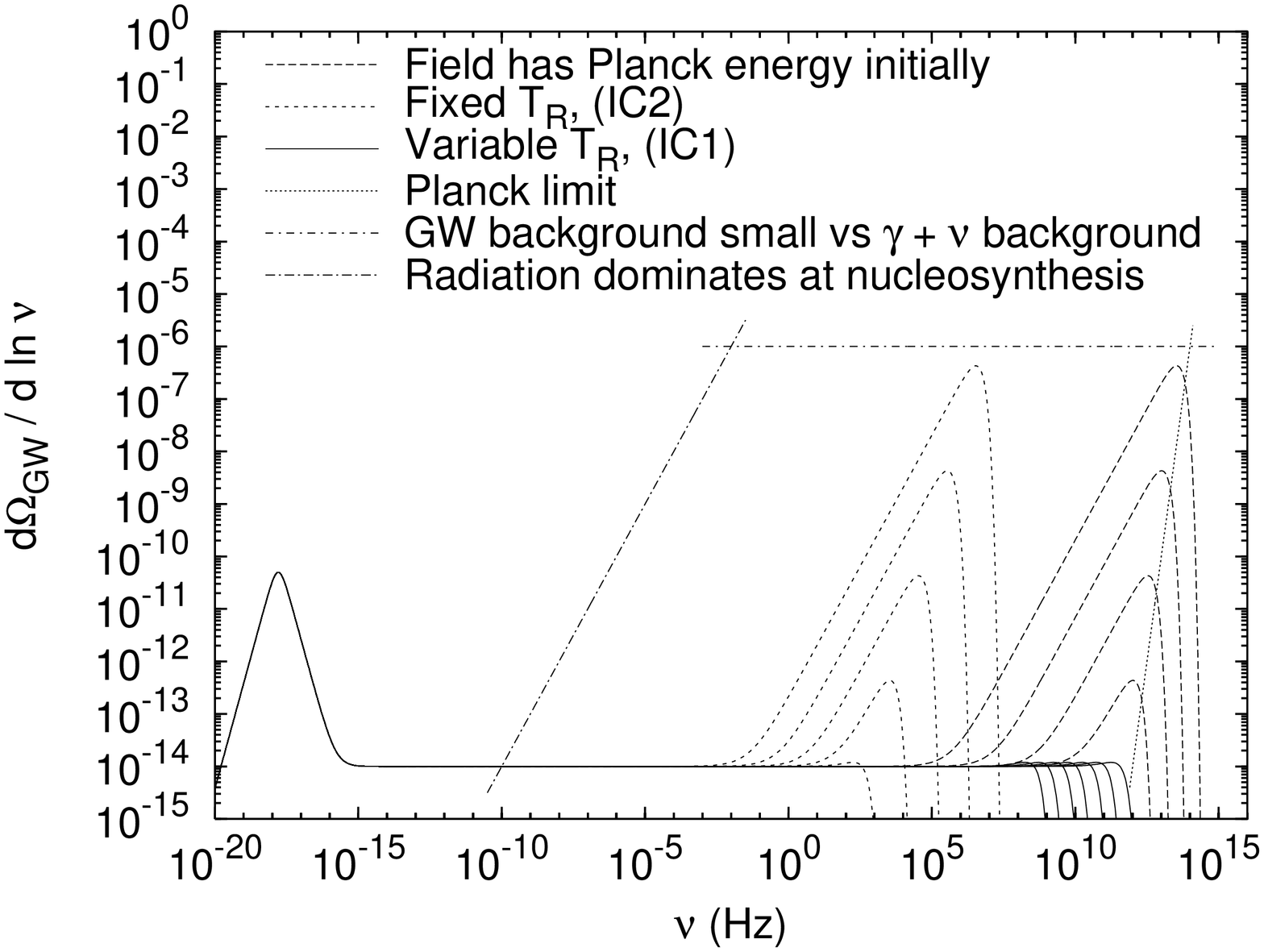,width=3.5in,angle=270}
\psfig{figure=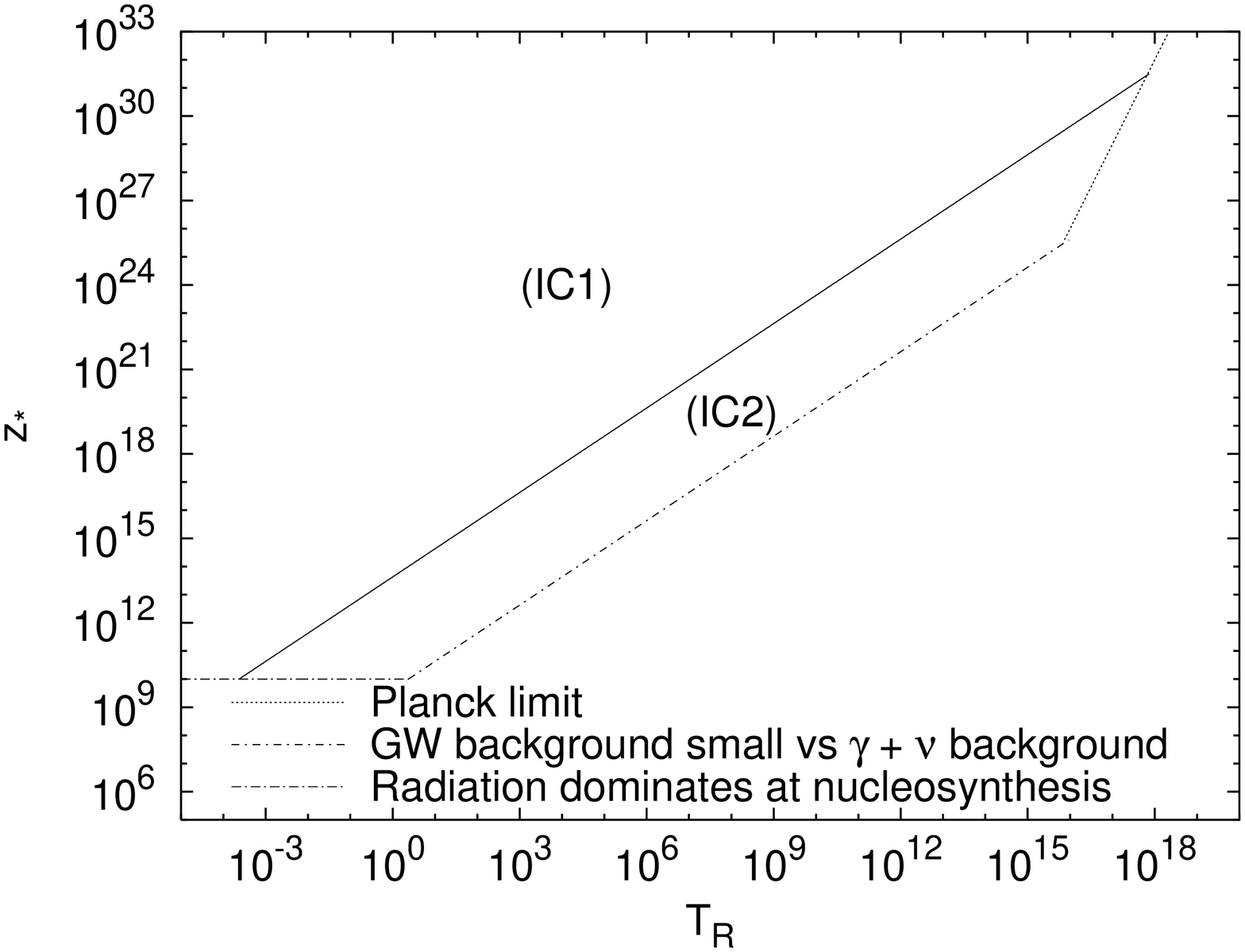,width=3.5in,angle=270}
}
\caption{Variation of the gravitational waves spectrum with the cut-off
and with the epoch of transition between scalar field in kinetic regime
and radiation [left] and contour plot [right] of the safe zone for
nucleosynthesis for the parameter set $(T_R,z_*)$ (above the dotted and
dot-dashed lines). Both plots are for a spectral index $n_T=0$ and for
maximum $A_T$ allowed by COBE measurements. The solid line is obtained
by imposing that the scalar field dominates at the end of inflation:
points lying on or above this line have a corresponding solid power
spectrum on the left plot. The horizontal dot-dash line is obtained by
imposing that we are radiation dominated at nucleosynthesis
(\IE{} $z_*>10^{10}$). The diagonal dotted line is obtained by imposing
that the energy density at the end of inflation is smaller than the
Planck density. The diagonal dot-dashed line (``GW background\ldots'')
is obtained by imposing that $\Omega_\GW^0<10^{-6}$.}
\label{reheating}
\end{figure}

To finish, let us discuss the total energy density of gravitational
waves $\Omega_\GW^0$ today. As pointed out in~\cite{giovannini99}, it
has also to be negligible at nucleosynthesis; this constraint is more
drastic than the only requirement that $\Omega_\phi^0$ be negligible
at that time. Let us emphasize that the constraint on $\Omega_\phi^0$
cannot be avoided (since it involves background dynamics) whereas the
one on $\Omega_\GW^0$ depends on $A_T$ and $n_T$ and thus leads to a
combined constraint on the initial conditions of the scalar field and
on the initial power spectrum of the gravitational waves. Besides
$A_T$ and $n_T$, $\Omega_\GW^0$ mainly depends on the initial values
of $\rho_\phi$ and $\rho_{\rm rad}$ which can be parametrised by the
reheating temperature $T_R$ (related roughly to $\rho_{\rm rad}$ at
that time) and the redshift $z_*$ of equality between the kinetic
scalar field era and the radiation era (related roughly to
$\rho_\phi/\rho_{\rm rad}$ at the end of reheating). $\Omega_\GW^0$
can be estimated by the surface of the spectrum below the part with a
positive slope (\IE{} the high frequency part; see \FIG{\ref{plot9}})
and thus of order
\begin{equation}
\Omega_\GW^0\sim\frac{k_R}{k_*}\left.
\frac{\ddd\Omega_{\rm GW}}{\ddd\ln k}\right|_{k_*}
\end{equation}
where $k_R$ and $k_*$ are respectively the modes entering the Hubble
radius at the reheating and at $z_*$. Thus the ``bump'' at short
wavelength cannot be too high. Moreover, the energy density at the end
of reheating cannot be higher that Planck scale, so that it fixes a
limit on the shortest mode in which gravitational waves are
produced. On \FIG{\ref{reheating}}, we first plot [left] the variation
of the gravitational wave spectrum with the parameters $(T_R,z_*)$ and
we then give [right] the ``safe'' zone of parameters for
nucleosynthesis [for $n_T=0$] and defined~\cite{giovannini99,vilenkin}
by $\Omega_{\rm GW}^0\lesssim10^{-6}$. Let us briefly explain how this
bounds are obtained.
\begin{enumerate}

\item We first rephrase in terms of $T_R$ the fact that the field
is dominating at the end of the inflation phase, \IE
\begin{equation}
z_*<z_R\Longleftrightarrow z_*<\alpha_1T_R,
\end{equation}
where $\alpha_1$ is some numerical coefficient. This corresponds to the
solid line on \FIG{\ref{reheating}}.

\item We then impose that the scalar field is subdominant at
nucleosynthesis, \IE{} that
\begin{equation}
z_* > 10^{10}.
\end{equation}
This corresponds to the horizontal dash-dot line on
\FIG{\ref{reheating}}.

\item At the end of the inflation phase, we want the energy density to
be smaller that the Planck energy density. If the scalar field is
dominating if gives
\begin{equation}
\rho_\phi^0\frac{(1+z_R)^6}{(1+z_*)^2}(1+z_{\rm eq})<\rho_\PL
\Longleftrightarrow
z_*>\alpha_2T_R^3
\end{equation}
where $\alpha_2$ is another numerical coefficient. This corresponds to
the dot line on \FIG{\ref{reheating}}. Note that since we are in a
field dominated era $H^2\propto a^2$ and thus on this ``Planck limit''
we have $k_R \propto z_R^2 / z_* $ and $k_* \propto z_*$ (and thus
$k_R \propto 1/z_R$ and $k_* \propto z_R^3$) from wich we conclude that
the maximum of the power spectrum is roughly located on a curve
$(1/z_R,1/z_R^4)$ [see \FIG{\ref{reheating}}].

\item To finish, we want that the gravitational waves energy density
does not alter nucleosynthesis, \IE{} that
\begin{equation}
\Omega_\GW^0\lesssim10^{-6}
\Longleftrightarrow
T_R < \alpha_3 z_*
\end{equation}
where $\alpha_3$ is a third numerical coefficient. This corresponds
to the dot-dash line on \FIG{\ref{reheating}}.
\end{enumerate}

For all the points $(T_R,z_*)$ above the dotted and dot-dashed lines of
\FIG{\ref{reheating}} [right], there is no excess of gravitational
waves. The solid line separates the two sets of initial conditions zwe
have considered. We must emphasize that this result was obtained for
$n_T=0$ and that the spectrum can be tilted, which modifies the bounds
on the parameter set $(T_R,z_*)$ (more precisely, taking smaller $A_T$
or negative $n_T$ lowers the diagonal dot-dashed line). Such
constraints may be important for instance while considering models
where a scalar field dominates at baryogenesis~\cite{joyce}.

In the case of a ``blue'' initial power spectrum (\IE{} with $n_T > 0$
in our notations, or $\beta > -2$ in the notations
of~\cite{grish2000}), as advocated for example in~\cite{grish2000},
the flat branch of \FIG{\ref{plot9}} (corresponding to the
``semi-hard'' branch of~\cite{giovannini99b}) is already tilted,
giving as stronger constraint on our model. For instance, if $n_T =
0.2$, the amplitude at $\nu = 10 \, \rm{GHz}$ is boosted by a factor
$\simeq 3 \times 10^5$. As a consequence, the quantity of
gravitational waves at high frequency cannot be boosted as much as in
the case of a scale invariant spectrum, and the allowed range of
parameters for our model (see \FIG{\ref{reheating}}) is narrowed.

\section{Conclusion}
\label{par6}

In this article, we have studied some properties of quintessence
models with a non-minimally coupled scalar field among which the
spectrum of gravitational waves.

We have shown that such a quintessence field can behave as a fluid
with $\omega<-1$ and our models lead to $-3\lesssim \omega \lesssim 0$
when the field dominates. We related the energy scale $\Lambda$ of the
potential to its slope $\alpha$ and to the scalar field energy density
today $\Omega_\phi^0$. In particular, we showed that $\Lambda$ is
almost independent of $\Omega_\phi^0$. The {\it coincidence problem},
\IE{} the fact that $\Omega_\phi^0\sim1$ implies a tuning of $\Lambda$
(roughly the precision on $\Lambda$ has to one order of magnitude
higher than the one on $\Omega_\phi^0$) which is however less severe
than the fine tuning needed for a cosmological constant. This being
fixed, the tracking mechanism allows to span a very wide range of
initial conditions for the scalar field and there is no fine tuning in
that respect.

We then showed that the combined study of the gravitational waves
energy spectrum and of their imprint on the CMB radiation temperature
and polarisation enables to extract many complementary informations on
the models:
\begin{itemize}
\item the CMB mainly gives results on $\xi$, $\Omega_\phi^0$ and $n_T$,
\item the energy spectrum gives results on the initial conditions of the
scalar field.
\end{itemize}
As pointed out in~\cite{giovannini99,vilenkin}, there is an excess of
gravitational waves today if inflation ends by a kinetic phase. In
that case, one has to check that both $\Omega_\phi$ and $\Omega_{\rm
GW}$ are negligible at the time of nucleosynthesis and we relate the
amount of gravitational waves today to the reheating temperature and
the time of equality between the kinetic scalar era and the radiation
era.

We also pointed out that gravitational waves are damped by the
anisotropic stress of radiation, which implies that the CMB anisotropy
and polarisation spectra are lowered roughly by 10\% for high
multipoles. It was also shown that the amplitude of the gravitational
waves spectrum for inverse power law potentials is $\sim 30$\% higher
than for SUGRA-like potentials at high frequency. Indeed this is
probably not detectable by coming experiments but it could ultimately
lead to a signature of supergravity.

\acknowledgements

It is a pleasure to thank Pierre Bin\'etruy, Nathalie Deruelle,
Thibault Damour, David Langlois, Patrick Peter and Filipo Vernizzi for
fruitful discussions.



\end{document}